\documentclass[fleqn,12pt]{article}
\usepackage{amsfonts,epsfig,latexsym}

\newcommand{\Ga}{\alpha}

\newcommand{\GD}{\Delta}
\newcommand{\Gd}{\delta}
\newcommand{\Ge}{\epsilon}
\newcommand{\Geps}{\varepsilon}
\newcommand{\Gg}{\gamma}
\newcommand{\GG}{\Gamma}

\newcommand{\GL}{\Lambda}

\newcommand{\Go}{\omega}

\newcommand{\GTh}{\Theta}

\newcommand{\CB}{{\cal B}}

\newcommand{\CD}{{\cal D}}

\newcommand{\CL}{{\cal L}}
\newcommand{\CP}{{\cal P}}
\newcommand{\CQ}{{\cal Q}}

\newcommand{\CO}{{\cal O}}

\newcommand{\CU}{{\cal U}}
\newcommand{\CV}{{\cal V}}

\newcommand{\dA}{{\dot{A}}}
\newcommand{\dB}{{\dot{B}}}

\newcommand{\Bchi}{\overline{\chi}}
\newcommand{\Bpsi}{\overline{\psi}}

\newcommand{\ft}[2]{{\textstyle {\frac{#1}{#2}} }}

\newcommand{\dd}{\partial}
\newcommand{\tr}{{\rm tr \,}}
\newcommand{\diag}{{\rm diag \,}}

\newcommand{\ra}{\rightarrow}
\newcommand{\I}{{ i}}

\newcommand{\be}{\begin{equation}}
\newcommand{\ee}{\end{equation}}
\newcommand{\ben}{\begin{displaymath}}
\newcommand{\een}{\end{displaymath}}
\newcommand{\ba}{\begin{eqnarray}}
\newcommand{\ea}{\end{eqnarray}}
\newcommand{\nn}{\nonumber}
\newcommand{\non}{\nonumber\\}
\newcommand{\bean}{\begin{eqnarray*}}
\newcommand{\eean}{\end{eqnarray*}}

\newcommand{\mathon}{\mathversion{bold}}
\newcommand{\mathoff}{\mathversion{normal}}

\newcommand{\la}{\label}
\newcommand{\ci}{\cite}

\newcommand{\Ref}[1]{(\ref{#1})}

\makeatletter
\@addtoreset{equation}{section}
\makeatother

\newcommand{\vl}{{\vphantom{[}}}

\newcommand{\cro}{\!\times\!}
\newcommand{\equ}{\!=\!}
\newcommand{\pls}{\!+\!}
\newcommand{\mis}{\!-\!}

\newcommand{\Si}{{\bf i}}
\newcommand{\Sj}{{\bf j}}
\newcommand{\SBi}{{\bf j}}
\newcommand{\sO}{{\mathcal O}}

\newcommand{\Mat}{{\mathcal S}}
\newcommand{\adsr}{{L}}

\newcommand{\acq}{{\sf q}}

\begin{document}

\thispagestyle{empty}

\begin{flushright}
ROM2F/2001/40 \\
SPIN-2001/31\\
{\tt hep-th/0112154}
\end{flushright}


\begin{center}
{\bf\Large An Exact Holographic RG Flow}
\medskip

{\bf\Large Between 2d Conformal Fixed Points}

\bigskip\bigskip

{\bf Marcus~Berg\footnote{\tt berg@roma2.infn.it} and
Henning~Samtleben\footnote{\tt H.Samtleben@phys.uu.nl}}

\vspace{.3cm}  
$^1${Department of Physics,
University of Rome, Tor Vergata \\
Via della Ricerca Scientifica, 00133 Rome, Italy}

\vspace{.5cm}

$^2${Spinoza Insituut, Universiteit Utrecht, Leuvenlaan 4\\
Postbus 80.195, 3508 TD Utrecht, The Netherlands}

\end{center}
\renewcommand{\thefootnote}{\arabic{footnote}}
\setcounter{footnote}{0}
\bigskip
\medskip
\begin{abstract}

We describe a supersymmetric RG flow between conformal fixed points of
a two-dimensional quantum field theory as an analytic domain wall
solution of the three-dimensional $SO(4)\cro SO(4)$ gauged
supergravity. Its ultraviolet fixed point is an $N\equ(4,4)$
superconformal field theory related, through the double D1-D5 system,
to theories modeling the statistical mechanics of black holes. The
flow is driven by a relevant operator of conformal dimension
$\Delta\equ\ft32$ which breaks conformal symmetry and breaks
supersymmetry down to $N\equ(1,1)$, and sends the theory to an
infrared conformal fixed point with central charge $c_{\rm IR} =
c_{\rm UV} / 2$.

Using the supergravity description, we compute counterterms, one-point
functions and fluctuation equations for inert scalars and vector
fields, providing the complete framework to compute two-point
correlation functions of the corresponding operators throughout the
flow in the two-dimensional quantum field theory.  This produces a toy
model for flows of $N=4$ super Yang-Mills theory in 3+1 dimensions,
where conformal-to-conformal flows have resisted analytical solution.

\end{abstract}

\renewcommand{\thefootnote}{\arabic{footnote}}
\vfill
\leftline{{ December 2001}}

\setcounter{footnote}{0}
\newpage

\section{Introduction}

In the last few years, the interest in the connection between gauge
theories and string theory has been re-ignited by the AdS/CFT
correspondence~\cite{Mald98,GuKlPo98,Witt98}.  Extending the success
of the correspondence to theories that flow to some conformal field
theory at a fixed point is of great interest; this is accomplished by
holographic renormalization group methods
~\cite{BalKra99b,dBVeVe00,dHSoSk01,BiFrSk01,BiFrSk01a}.  Using this
extension of the correspondence, we describe a supersymmetric
renormalization group (RG) flow between conformal fixed points of a
two-dimensional quantum field theory.  To the best of our knowledge,
this is the first example of a holographic flow between conformal
fixed points which is exact, i.e.\ an explicit solution for the bulk
spacetime. The boundary theory is a large (or ``double'') $N\equ(4,4)$
superconformal field theory~\cite{BoPeSk98,EFGT99,dBPaSk99}, in which
the flow is driven by a relevant operator of conformal dimension
$\Delta\equ\ft32$, which we interpret as a mass term for chiral
superfields, in close analogy with the flow of~\cite{FGPW99}.  This
$\Delta\equ\ft32$ operator breaks conformal symmetry, and it breaks
supersymmetry to $N\equ(1,1)$.  Our general interest in these theories
is twofold: first, this type of CFT is known to supply quantitative
understanding of black hole quantum mechanics.  Second, our flow can
serve as a toy model for flows of physically relevant theories, a
prominent example being flows of $N\equ1$ super Yang-Mills (SYM)
theory in $3+1$ dimensions.  In the rest of the introduction, we
expand on these two motivations and give an overview of the new
results in this paper.

The above-mentioned $N\equ(4,4)$ conformal field theory is related to
theories modeling the statistical mechanics of black holes.  An
important example is the D1-D5 system on a torus $T^4$, which has a
complementary supergravity description as a five-dimensional black
hole with three charges.  In~\cite{MaMoSt99}, the $U(1)^4$ symmetry of
translations in the 4-torus was seen as a limit of an affine $SU(2)
\times SU(2) \times U(1)$ symmetry where the ratio $\alpha=k^+/k^-$ of
$SU(2)$ levels becomes small.  This larger symmetry was used to
compute the density of states in the D1-D5 system on $T^4$ despite the
vanishing of the elliptic genus in this case.  Now, the affine $SU(2)
\times SU(2) \times U(1)$ is part of the large $N\equ4$ superconformal
algebra~${\mathcal A}_{\gamma}$~\cite{SeTrVP88}.\footnote{We use the
standard notation $k^+$, $k^-$, for the levels of the two affine
$SU(2)$ factors, while $\gamma$ is given by $\gamma=k^+/(k^+\pls
k^-)=\Ga/(1\pls\Ga)$. The central charge of ${\mathcal A}_{\gamma}$ is
$c=6k^+k^-/(k^+\pls k^-)$.} (As a reminder, the small $N\equ4$ algebra
only has one affine $SU(2)$).

In terms of branes, these large symmetries correspond to a ``double''
D1-D5 system \cite{BoPeSk98,dBPaSk99}, from which the standard D1-D5
system is recovered in the limit $\Ga\ra0$ where the charges of one of
the systems are much greater than the charges of the other. Close to
this limit, the double D1-D5 system is an interesting (and rather
puzzling) ten-dimensional deformation of the physically relevant D1-D5
system, even though its supergravity description might not be a
deformation of a five-dimensional black hole in a five-dimensional
sense. It remains somewhat mysterious; as emphasized
in~\cite{dBPaSk99}, strings stretching between the two D5-branes
induce nonlocal couplings between the worldvolume theories.

At low energy, the double D1-D5 system yields an $AdS_3 \times S^3
\times S^3 \times S^1$ geometry admitting 16 Killing spinors, i.e.\
which is half maximally supersymmetric. The isometries of this
geometry form two copies of the supergroup $D^1(2,1;\alpha)$, where
the ratio of brane charges $\alpha$ now coincides with the ratio of
the two sphere radii. This supergroup shares the bosonic subgroup
$SU(2)^2 \times SU(1,1)$ with ${\mathcal A}_{\gamma}$.  In this paper,
we mostly concentrate on the case of $\alpha=1$, when the isometry is
two copies of $D^1(2,1;1)=OSp(4|2)$.

{}From a black hole point of view, it is interesting to try to
understand what remains of the relation to the black hole picture when
less supersymmetry is present and conformal symmetry of the
worldvolume theory is broken.  Adding a Lorentz-invariant relevant
operator to the Lagrangian accomplishes this; in particular for
operators that completely break supersymmetry, one could imagine an RG
flow that describes temperature effects in the black hole.  At the
moment, we have nothing more to add about this --- of course, without
supersymmetry it is not clear how to control corrections in the
transition to the supergravity regime.

A general reason to study holographic RG flows in two dimensions is
simply to provide toy models for flows of strongly coupled gauge
theories in higher dimensions, among which $N=1$ SYM is the example of
greatest phenomenological interest.  In two dimensions, one can study
features of the correspondence that are under much better control than
in four dimensions.  For instance, much information is encapsulated in
the central charge $c$.  The Zamolodchikov $c$-theorem \cite{Zamo86}
states that $c$ is smaller at an infrared fixed point, i.e.\ that
degrees of freedom associated to massive fields become unimportant at
distances longer than the Compton wavelength of those massive
excitations.  Instead of holographically proving the $c$-theorem as in
four dimensions, we have the luxury of a well-established result on
the CFT side.  Indeed, all fixed points we find satisfy the
Zamolodchikov $c$-theorem.

Our flow can be viewed as a two-dimensional analogue of the
supersymmetric Freedman-Gubser-Pilch-Warner (FGPW) flow between $N=4$
SYM and an $N=1$ superconformal field theory in 3+1 dimensions,
obtained by giving mass to one of the three chiral superfields
\cite{FGPW99}. (In contrast, the well-studied $N=1^*$ theory
flow \cite{GPPZ00,PolStr00} corresponds to the three chiral
superfields receiving equal mass.)  In four dimensions, the spacetime
holographically dual to this flow could only be described numerically.
This is a great drawback as one would need an exact flow to be able to
compute correlation functions.  While there is no obvious reason why
the flow equations themselves (which are always one-dimensional)
should be simpler in three-dimensional spacetime than in five, we find
an analytic solution to the three-dimensional Killing spinor
conditions.

The tool of holographic renormalization is currently gauged
supergravity. In the FGPW flow of $N=4$ SYM in 3+1 dimensions, the
G\"unaydin-Romans-Warner five-dimensional gauged supergravity, with an
$E_{6(6)}/USp(8)$ scalar manifold \cite{GuRoWa86}, provided the proper
setting.  Although the full nonlinear dimensional reduction of
ten-dimensional supergravity on $AdS_5 \times S^5$ is not known, the
aforementioned five-dimensional gauged supergravity has been argued to
be a consistent truncation describing states of lowest mass in this
reduction.  The analogous framework for the holographic dual of a 1+1
field theory would be found among three-dimensional gauged
supergravities; these theories were constructed in
\cite{NicSam01,NicSam01b}. In these papers,
consistency conditions were condensed to a group theory condition from
which a menu of allowed gauge groups could be compiled. Returning to
the case of half-maximal supersymmetry, the gauge groups are
particular subgroups~\cite{NicSam01b} of the isometries of the scalar
manifold $SO(8,n)/(SO(8)\cro SO(n))$. Since we are interested in
compactifications where the internal geometry includes an $S^3 \times
S^3$, the theory relevant for our analysis has local symmetry $SO(4)
\cro SO(4)$ with two independent coupling constants, corresponding to
the isometry groups and radii of the two three-spheres,
respectively. As we shall discuss in more detail below, this theory
indeed reproduces the spectrum of lowest mass states found in the
reduction of supergravity on $AdS_3\cro S^3\cro S^3\cro S^1$.

We should also emphasize that the undeformed boundary theory has some
mysterious properties that we do not attempt to clarify in this paper.
Unlike in the case of $N=2$ and small $N=4$ theories, the BPS relation
between conformal dimension and charge in the large $N=4$ theory is
nonlinear, see equation (\ref{hlu}).  As pointed out in
\cite{dBPaSk99}, the nonlinear contribution is subleading in $1/N$,
hence the BPS mass formula receives string loop corrections.  In
supergravity, this nonlinear part seems invisible (contrast (\ref{hl})
and (\ref{hlu})), which makes it difficult to establish a precise
correspondence including multiparticle states for general $\alpha$.
Nevertheless, in \cite{EFGT99}, a correspondence was proposed for
rational values of the ratio $\alpha$.

The paper is organized as follows. In section~2 we review the
three-dimensional $N\!=\!8$ gauged supergravity with gauge group
$SO(4)\cro SO(4)$, describing the lowest mass states in the reduction
of supergravity on $AdS_3\cro S^3\cro S^3\cro S^1$. We study the
scalar potential in certain truncations and find several stable
extremal points. In particular, we uncover two extrema which preserve
$N=(1,1)$ supersymmetry. The spectrum of physical fields around these
extrema is given in section~2.3 and the appendix. In section~3 we
present an analytic kink solution for the metric and the scalar fields
which interpolates between the central maximum of the potential and
one of the supersymmetric extrema. Interpreted as a holographic
renormalization group flow, this solution flows between conformal
fixed points with $c_{\rm IR}/c_{\rm UV} = 1/2$. We moreover find that
all the operators in the IR theory have rational conformal dimension.

Section~4 contains the computation of counterterms for inert scalars
and the 1-point functions of their CFT duals. In section~5 we derive
the linearized fluctuation equations for inert scalars and vector
fields around our flow solution. We show that they may all be reduced
to two second-order ``universal'' differential equations.  Together
with the 1-point functions, the properly normalized solutions to
these equations encode the entire information about 2-point
correlation functions of the dual operators in the boundary
theory. The appendix contains an explicit parametrization of the
scalar potential of the gauged supergravity, and a collection of
stable extrema together with their spectra.

\mathon
\section{$D\equ3$, $N\equ8$ supergravity with local $SO(4)\cro SO(4)$}
\mathoff

\subsection{Lagrangian}

As discussed in the introduction, we are interested in
compactifications where the internal geometry includes an $S^3 \times
S^3$, leading to an $SO(4) \times SO(4)$ gauge symmetry in the
three-dimensional effective theory, with 16 real supercharges.  This
theory is the three-dimensional $N\equ8$ gauged supergravity with
local $SO(4)\cro SO(4)$ symmetry. Its matter sector consists of $n$
multiplets each containing $8$ scalars and $8$ fermions, whereas
graviton, gravitini, and the $12$ vector fields are non-propagating in
three dimensions, see~\Ref{LCS} and the subsequent discussion.  An
important difference to the maximally supersymmetric case is that from
a three-dimensional point of view, we may turn on any number $n$ of
matter multiplets in the classical supergravity, although anomaly
cancellation can constrain the value of $n$ in the quantum theory. (In
a three-dimensional bulk, there is no chiral anomaly, so it would have
to come from the boundaries.)  The $8n$ scalars parametrize the coset
manifold $SO(8,n)/(SO(8)\cro SO(n))$.

The Lagrangian of this theory is given by \cite{NicSam01b}
\ba
\CL&=& -\ft14 \,\sqrt{G}R
+ \CL_{CS} + 
\ft14\,\sqrt{G} G^{\mu\nu}\,\CP_\mu^{Ir} \CP_\nu^{\,Ir} 
- \sqrt{G}\,V + \CL_F\;,
\la{L}
\ea
where $\CL_F$ contains the fermionic terms, explicitly given in
\Ref{LF} below, and $G_{\mu\nu}$ is the bulk metric.
We use signature $(+--)$. 
Indices $I, J, \dots$ and indices $r, s, \dots$ denote
the vector representations of $SO(8)$ and $SO(n)$, respectively. The
12 vector fields transform in the adjoint representation of the gauge
group
\be
SO(4)^+\times SO(4)^- ~\subset~ SO(8) ~\subset~ SO(8,n)\;,
\la{gaugeG}
\ee
where we use $+$ and $-$ superscripts to distinguish the two
three-spheres.  The vector fields are collectively denoted by
$B_\mu{}^{IJ}=B_\mu{}^{[IJ]}$, for $I, J \in \{1,2,3,4\}$ or $I, J \in
\{5,6,7,8\}$, respectively, corresponding to the two factors in
\Ref{gaugeG}. In contrast to higher-dimensional gauged supergravities,
the dynamics of the vector fields is governed by a Chern-Simons term
\be
\CL_{CS} ~=~
-\ft14\,\Ge^{\mu\nu\rho}\,g\GTh_{IJ,KL}\,B_\mu{}^{IJ}
\Big(\dd_\nu B_\rho\,{}^{KL}
+\ft83\,g\, \eta^{KM}\,\GTh_{MN,PQ}\, 
B_\nu{}^{PQ} B_\rho{}^{NL} \Big) \;,
\la{LCS}
\ee
indicating that this dynamics is pure gauge, i.e.\ the vector fields
do not carry physical degrees of freedom.\footnote{Specifically, one
may view the vector fields as nonlocal functions of the scalar fields,
entirely defined by the first order duality equations induced
by~\Ref{LCS} up to local gauge freedom. Equivalently, one may fix
this gauge by eliminating some of the scalar fields, whereby the
vector-scalar duality equations become massive self-duality equations
for the vector fields. This is illustrated in the linearized analysis
in section~5. \label{footn}} Here, the parameter $g$ denotes the gauge
coupling constant, and the tensor $\GTh_{IJ,KL}$ describes the
embedding of the gauge group into $SO(8)$:\footnote{Our conventions
for the $\Ge$-symbols are
$\Ge^\vl_{1234}=\Ge^\vl_{5678}=\Ge^\vl_{12345678}=1$, while we use a
representation of $SO(8)$ $\Gamma$-matrices in which $\Gamma^{[4]}$ is
selfdual according to $\GG^{IJKL} =
\Ge^\vl_{IJKLMNPQ}\,\GG^{MNPQ}\;$.}
\be
\GTh_{IJ,KL} ~=~ \left\{
\begin{array}{rl}
\alpha \, \Ge^\vl_{IJKL} & \mbox{for}\quad  I,J,K,L \in \{1,2,3,4\} \\
\Ge^\vl_{IJKL} & \mbox{for} \quad I,J,K,L \in \{5,6,7,8\} \\[.5ex]
0 & \mbox{otherwise}
\end{array}
\right. \;.
\la{theta}
\ee
The free parameter $\Ga$ describes the ratio of coupling constants of
the two $SO(4)$ factors~\Ref{gaugeG}, alias the ratio of radii of the
two three-spheres or the ratio of charges of the two D1-D5 systems,
cf.\ the discussion in the introduction. The scalar sector in \Ref{L}
is parametrized by $SO(8,n)$ matrices $\Mat$ which define the currents
$\CP_\mu^{Ir}$ according to
\be
\Mat^{-1}\CD_\mu \Mat ~\equiv~  
\Mat^{-1} (\dd_\mu +   
g \GTh_{IJ,KL} \,B_\mu{}^{IJ} X^{KL})\, \Mat ~\equiv~ 
\ft12\CQ_\mu^{IJ} X^{IJ} 
+ \ft12\CQ_\mu^{rs} X^{rs} + \CP_\mu^{Ir} Y^{Ir} \;,
\la{current}
\ee
where $X^{IJ}$, $X^{rs}$ denote the compact generators of $SO(8,n)$, 
and $Y^{Ir}$ the noncompact ones. The scalar potential
is given by 
\be
V ~=~-\ft14\,g^2\,
\left(A_1^{AB}A_1^{AB}-\ft12\,A_2^{A\dA r}A_2^{A\dA r}\right) \;,
\la{potential}
\ee
in terms of the $SO(8)$ tensors $A_1$, $A_2$
\ba
A_{1\, AB}&=& 
-\ft1{48}\,\GG^{IJKL}_{AB}\,
\CV^{MN}{}_{IJ}\,\CV^{PQ}{}_{KL}\,\GTh_{MN,PQ}
\;,
\non
A_{2\, A\dA r}&=&
-\ft1{12}\,\GG^{IJK}_{A\dA}\,
\CV^{MN}{}_{IJ}\,\CV^{PQ}{}_{Kr}\,\GTh_{MN,PQ} \;,
\nn
\ea
with the scalar matrix $\CV$ obtained from expanding
\ben
\Mat^{-1} X^{IJ} \Mat ~\equiv~ 
\ft12\,\CV^{IJ}{}_{MN}\,X^{MN}\, + 
\ft12\,\CV^{IJ}{}_{rs}\,X^{rs}\, + \CV^{IJ}{}_{Kr}\,Y^{Kr}
\;.
\een
The Lagrangian~\Ref{L} has a local $SO(8)\times SO(n)$ symmetry
(corresponding to the redundancies of the coset structure
$SO(8,n)/(SO(8)\cro SO(n))$), which acts by right multiplication on
$\Mat$. The local $SO(4)\cro SO(4)$ gauge symmetry acts by left
multiplication on $\Mat$. In addition, \Ref{L} possesses a remaining
global $SO(n)$ symmetry which likewise acts by left multiplication on
$\Mat$, rotating the $n$ matter multiplets.

The potential~\Ref{potential} has a local maximum at $\Mat=I_{8n}$,
which yields an $N\equ(4,4)$ supersymmetric $AdS_3$ solution of
\Ref{L}. The value of the potential at this point is
\be
V_0 ~\equiv~ V|^\vl_{\Mat=I_{8n}} ~=~ -8g^2\,  (1+\Ga)^2 \;,
\la{V00}
\ee
i.e.\ the three-dimensional AdS radius $\adsr_0$ is related to the
gauge coupling constant $g$ by
\be
\adsr_0=\frac1{4|(1\!+\!\Ga) g|} \;,\qquad\mbox{with}\quad
R_{\mu\nu} = -4g^2 V_0G_{\mu\nu} = 
\frac2{\adsr_0^2}\, G_{\mu\nu} \;.
\la{AdSrad}
\ee
We will later talk about AdS spaces with other radii $L$, but $L_0$
will always denote the radius at $\Mat=I_{8n}$, which should be set to
unity to obtain correctly normalized CFT correlators (see section 2 of
\cite{BiFrSk01} for a careful discussion of this point). Note that an
$S^3\cro S^3$ compactification corresponds to positive values of $\Ga$
--- as can be seen from (\ref{AdSrad}), the theory with $\alpha\equ-1$
admits a Minkowski solution, which we shall not further discuss here.
The $AdS_3$ supersymmetric solution has background isometry group
\be
D^1(2,1;\Ga)_L \times D^1(2,1;\Ga)_R \;,
\la{DD}
\ee
an $N=(4,4)$ superextension of the three-dimensional AdS group
$SU(1,1)_L\cro SU(1,1)_R$, with the parameter $\Ga$ from
\Ref{theta}. The spectrum around this local maximum may be organized
in representations of \Ref{DD}. Following \cite{dBPaSk99}, we denote a
short supermultiplet of \Ref{DD} by
$(\ell^+_L,\ell_L^-;\ell_R^+,\ell_R^-)_S$, where the
$\ell_{L,R}^{\pm}$ refer to the $SU(2)$ quantum numbers of the highest
weight state in the multiplet under the following bosonic subgroup of
\Ref{DD}:
\be
\left(SU(2)^+_L\times SU(2)^-_L\right) 
\times \left(SU(2)^+_R\times SU(2)^-_R\right) \;.
\la{G0}
\ee
The highest weight state state saturates the bound due to unitarity
\be
h \geq \gamma \ell_L^- + (1-\gamma)\ell^+_L
\label{hl}
\ee
where $h$ is the conformal dimension~$h_L$. In this notation, the
matter multiplets of \Ref{L} take the form $(\ft12,0;\ft12,0)_S$ or
$(0,\ft12;0,\ft12)_S$, each containing 8 scalar and 8 fermionic
fields. Their masses may be computed from linearizing the
Lagrangian~\Ref{L} around the local maximum $\Mat=I_{8n}$. Metric,
gravitini and vector fields are assembled into the ``nonpropagating
supermultiplets'' $(\ft12,\ft12;0,0)_S$ and $(0,0;\ft12,\ft12)_S$. The
complete list of the appearing supermultiplets, their decomposition
into states under \Ref{G0}, their masses in \Ref{L} and the conformal
weights $(h,\bar{h})$ under the $AdS_3$ part of \Ref{DD} are collected
in table~\ref{specASS}. We used the standard relations (\ref{MDsc}),
(\ref{MDve}) to find $\Delta=h+\bar{h}$ from $m^2 L_0^2$.
\begin{table}[htb]
\centering
\begin{tabular}{|c||c|c|c|c|c|} \hline
$(\ell^+_L,\ell_L^-;\ell_R^+,\ell_R^-)_S$&
Fields &$(\ell^+_L,\ell_R^+,\ell_L^-,\ell_R^-)$ & 
$(h,\bar{h})$ & $\GD$& $m^2\adsr_0^2$  \\
\hline\hline
$(\ft12,\ft12;0,0)_S$&graviton&$ (0,0,0,0)$ & 
$\big( 2, 0 \big)$ &
$\GD = 2$ & $0$ \\
\cline{2-6}
&gravitini&$ (\frac12,0,\frac12,0)$ & 
$\big( \frac32 , 0\big)$ &
$\GD = \frac32$ & $\ft14$\\
\cline{2-6}
&vectors&$ (1,0,0,0)$ & 
$\big( 1, 0 \big)$ &
$\GD = 1$ & $0$\\
\cline{3-6}
&&$ (0,0,1,0)$ & 
$\big( 1, 0 \big)$ &
$\GD = 1$ & $0$\\
\cline{2-6}
\hline\hline
$(0,0;\ft12,\ft12)_S$&graviton&$ (0,0,0,0)$ & 
$\big( 0, 2 \big)$ &
$\GD = 2$ & $0$\\
\cline{2-6}
&gravitini&$ (0,\frac12,0,\frac12)$ & 
$\big( 0, \frac32 \big)$ &
$\GD = \frac32$ & $\ft14$\\
\cline{2-6}
&vectors&$ (0,1,0,0)$ & 
$\big( 0,1 \big)$ &
$\GD = 1$ & $0$\\
\cline{3-6}
&&$ (0,0,0,1)$ & 
$\big( 0,1 \big)$ &
$\GD = 1$ & $0$\\
\cline{2-6}
\hline\hline
$(0,\ft12;0,\ft12)_S$&scalars&$(\frac12,\frac12,0,0)$ & 
$\big( \frac{\Ga}{2(1\pls\Ga)}, \frac{\Ga}{2(1\pls\Ga)} \big)$ &
$\GD_- = 1-\frac{1}{1\pls\Ga}$ & $-\frac{\Ga(2\pls\Ga)}{(1\pls\Ga)^2}$\\
\cline{3-6}
&&$(0,0,\frac12,\frac12)$ & 
$( \frac{1\pls2\Ga}{2(1\pls\Ga)}, \frac{1\pls2\Ga}{2(1\pls\Ga)} )$ &
$\GD_+ = 1+\frac{\Ga}{1\pls\Ga}$ & $-\frac{1\pls2\Ga}{(1\pls\Ga)^2}$\\
\cline{2-6}
&fermions&$(\frac12,0,0,\frac12)$ & 
$( \frac{\Ga}{2(1\pls\Ga)}, \frac{1\pls2\Ga}{2(1\pls\Ga)} )$ &
$\GD = \frac{1\pls3\Ga}{2(1\pls\Ga)}$ & 
$\frac{(1\mis\Ga)^2}{4(1\pls\Ga)^2}$\\
\cline{3-6}
&&$(0,\frac12,\frac12,0)$ & 
$( \frac{1\pls2\Ga}{2(1\pls\Ga)},\frac{\Ga}{2(1\pls\Ga)} )$ &
$\GD = \frac{1\pls3\Ga}{2(1\pls\Ga)}$ & 
$\frac{(1\mis\Ga)^2}{4(1\pls\Ga)^2}$\\
\hline\hline
$(\ft12,0;\ft12,0)_S$&scalars&$(\frac12,\frac12,0,0)$ & 
$( \frac{2\pls\Ga}{2(1\pls\Ga)}, \frac{2\pls\Ga}{2(1\pls\Ga)} )$ &
$\GD_+ = 1+\frac{1}{1\pls\Ga}$ & $-\frac{\Ga(2\pls\Ga)}{(1\pls\Ga)^2}$\\
\cline{3-6}
&&$(0,0,\frac12,\frac12)$ & 
$( \frac{1}{2(1\pls\Ga)}, \frac{1}{2(1\pls\Ga)} )$ &
$\GD_- = 1-\frac{\Ga}{1\pls\Ga}$ & $-\frac{1\pls2\Ga}{(1\pls\Ga)^2}$\\
\cline{2-6}
&fermions&$(\frac12,0,0,\frac12)$ & 
$(\frac{2\pls\Ga}{2(1\pls\Ga)}, \frac{1}{2(1\pls\Ga)} )$ &
$\GD = \frac{3\pls\Ga}{2(1\pls\Ga)}$ & 
$\frac{(1\mis\Ga)^2}{4(1\pls\Ga)^2}$\\
\cline{3-6}
&&$(0,\frac12,\frac12,0)$ & 
$( \frac{1}{2(1\pls\Ga)} ,\frac{2\pls\Ga}{2(1\pls\Ga)})$ &
$\GD = \frac{3\pls\Ga}{2(1\pls\Ga)}$ & 
$\frac{(1\mis\Ga)^2}{4(1\pls\Ga)^2}$\\
\hline
\end{tabular}
\caption{\small Lowest multiplets in the spectrum on 
$AdS_3\times S^3 \times S^3$.}
\label{specASS} 
\end{table}

This exactly reproduces the spectrum of lowest mass states in the
reduction of nine-dimensional supergravity on $AdS_3\times S^3\times
S^3$, found in~\cite{dBPaSk99}.\footnote{ Note that our ordering of
$SU(2)$ quantum numbers (3rd column in table~\ref{specASS}) differs
from that of Ref.~\cite{dBPaSk99} (1st column) by a permutation.}
More precisely, the infinite sums given in~\cite{dBPaSk99} contain
among their lowest mass multiplets one $(\ft12,0;\ft12,0)_S$ and one
$(0,\ft12;0,\ft12)_S$, corresponding to the scalar spectrum of~\Ref{L}
with $n\equ2$ matter multiplets coupled.

\subsection{Truncation and extrema of the potential}

We now search for extremal points of the scalar potential
\Ref{potential}. The dimension of the scalar manifold is $8n$; 
it is convenient to begin by explicitly fixing all remaining
symmetries of the Lagrangian \Ref{L}. These symmetries consist of a
local $SO(4)\cro SO(4)$ symmetry and a global $SO(n)$ symmetry
rotating the matter multiplets. For $n\equ4$ matter multiplets, say,
the scalar potential actually only depends on $8\cdot4\!-\!3\cdot6=14$
parameters out of the original 32. We have computed this scalar
potential analytically, and --- with some numerical help --- found
some of its extremal points, see appendix~\ref{AppExt} for a
collection of results. These results indicate that a minimal number of
$n\equ3$ matter multiplets is required for the potential to exhibit
nontrivial extremal points.

For our main examples, we will further truncate the theory. Following
the standard argument~\cite{Warn83}, extremal points found in the
truncation of the scalar sector to singlets under a subgroup of the
symmetry group may consistently be lifted to extrema of the full
theory. We are mainly interested in extrema which preserve part of the
$N\equ(4,4)$ supersymmetry. The representation content of the scalar
sector (table~\ref{specASS}) shows that there are no singlets under
any nontrivial product of subgroups of the left and right $R$-symmetry
groups $(SU(2)^+\times SU(2)^-)_{L,R}$ in \Ref{G0}. Thus, at most
$N\equ(1,1)$ supersymmetry can be preserved upon switching on scalar
fields.

Assuming $n\ge4$, we consider the following subgroup 
\ba
G_{\rm inv} ~\equiv~ SU(2)_{\rm inv}\times SO(n\mis4) &\subset& 
\left(SO(4)^+ \times SO(4)^- \right) \; 
\times \left( SO(4)^{\vphantom +} \times SO(n\mis4) \right) 
\non
&\subset& 
\left(SO(4)^+ \times SO(4)^- \right) \times SO(n) \;,
\la{Ginv}
\ea
of the global invariance group of the potential \Ref{potential}. The
$SU(2)_{\rm inv}$ factor in $G_{\rm inv}$ is embedded as the diagonal
of the six $SU(2)$ factors on the right hand side. We find that
under $G_{\rm inv}$ the scalar spectrum decomposes as
\be
4 \cdot {\bf (1,1)} + 6 \cdot {\bf (3,1)} + 2 \cdot {\bf (5,1)} +
2 \cdot {\bf (1,n\mis4)} + 2 \cdot {\bf (3,n\mis4)} \;.
\ee
Let us study the potential on the four-dimensional space of singlets
under $G_{\rm inv}$. This truncation corresponds to restricting the
scalar sector to $SO(8,n)$ matrices
\ba
\Mat &=&
\left( 
\begin{array}{cccc}
\cosh B & 0_{4\times(n-4)} & \sinh B &0_{4\times4}
\\
0_{(n-4)\times 4}&I_{n-4}&0_{(n-4)\times 4}&0_{(n-4)\times 4}\\
\cos A \sinh B& 0_{4\times (n-4)} &\cos A \cosh B& \sin A
\\
-\sin A \sinh B& 0_{4\times (n-4)} & -\sin A \cosh B& \cos A
\end{array}
\right) , \la{matL}\\[2ex]
&&{}
A=\diag(p_1, p_2, p_3, p_4 )\;,\quad
B=\diag(q_1, q_2, q_3, q_4 )
 \;,
\nn
\ea
and further setting $p_2\equ p_3\equ p_4$, $q_2\equ q_3\equ q_4$. As
another (mainly technical) simplification we will from now on set
$\Ga\equ1$. Substituting \Ref{matL} into \Ref{potential}, one obtains
the potential
\ba
g^{-2}\,V &=& -16 
- 4\, {\textstyle \sum_i}\; x_i^2 
- 4\, {\textstyle \sum_i }\; y_i^2 
+ 4\, {\textstyle \sum_{i<j<k}}\; x_i^2x_j^2x_k^2 
+ 4\, {\textstyle \sum_{i<j<k}}\; y_i^2y_j^2y_k^2
\non
&&{}
+ 8\, {\textstyle \prod_i}\; x_i^2 
+ 8\, {\textstyle \prod_i}\; y_i^2 
+16\,{\textstyle \prod_i}\; x_i^2 y_i^2 
-16\,{\textstyle \prod_i}\; \sqrt{1+x_i^2+y_i^2} \;,
\la{Vex}
\ea
where the indices $i, j, k$ in the sums and products run from 1 to 4,
and we have set
\be
x_i=\cos p_i \,\sinh q_i \;,\quad
y_i=\sin p_i \,\sinh q_i \;. 
\la{pqxy}
\ee
The kinetic term in \Ref{L} takes the form
\ba
\ft14 \sqrt{G} \, \CP_\mu^{Ir} \CP^\mu{}^{\,Ir} &=&
\frac{\sqrt{G}}4 \, \sum_{i=1}^4\;\Big(
\sinh^2\! q_i \, (\dd_\mu p_i)^2 
+ (\dd_\mu q_i)^2 \Big)
\non[1ex]
&=&
\frac{\sqrt{G}}4 \,\sum_{i=1}^4 
\frac{(1\!+\!y_i^2) (\dd_\mu x_i)^2 +(1\!+\!x_i^2) (\dd_\mu y_i)^2
- 2 x_i y_i (\dd_\mu x_i)(\dd^\mu y_i)}{1+x_i^2+y_i^2} \;.
\nn
\ea
Linearizing the scalar field equations around the origin gives rise to
\be
\Box\, x_i = \frac{3}{4 \adsr_0^2}\, x_i \;,\qquad
\Box\, y_i = \frac{3}{4 \adsr_0^2}\, y_i \;,
\la{lin34}
\ee
in agreement with table~\ref{specASS}. In particular, all scalar
fields satisfy the Breitenlohner-Freedman bound~\cite{BreFre82}, which
in these conventions is given by $m^2 \adsr^2 \ge -1$ for an AdS scale
of $L$. As an illustration of the scalar potential \Ref{Vex}, we have
depicted contour plots of particular slices in figures~1 and~2.
\bigskip
\bigskip

\hspace*{-5mm}
\begin{minipage}[htbp]{75mm}
\resizebox{70mm}{!}{\includegraphics{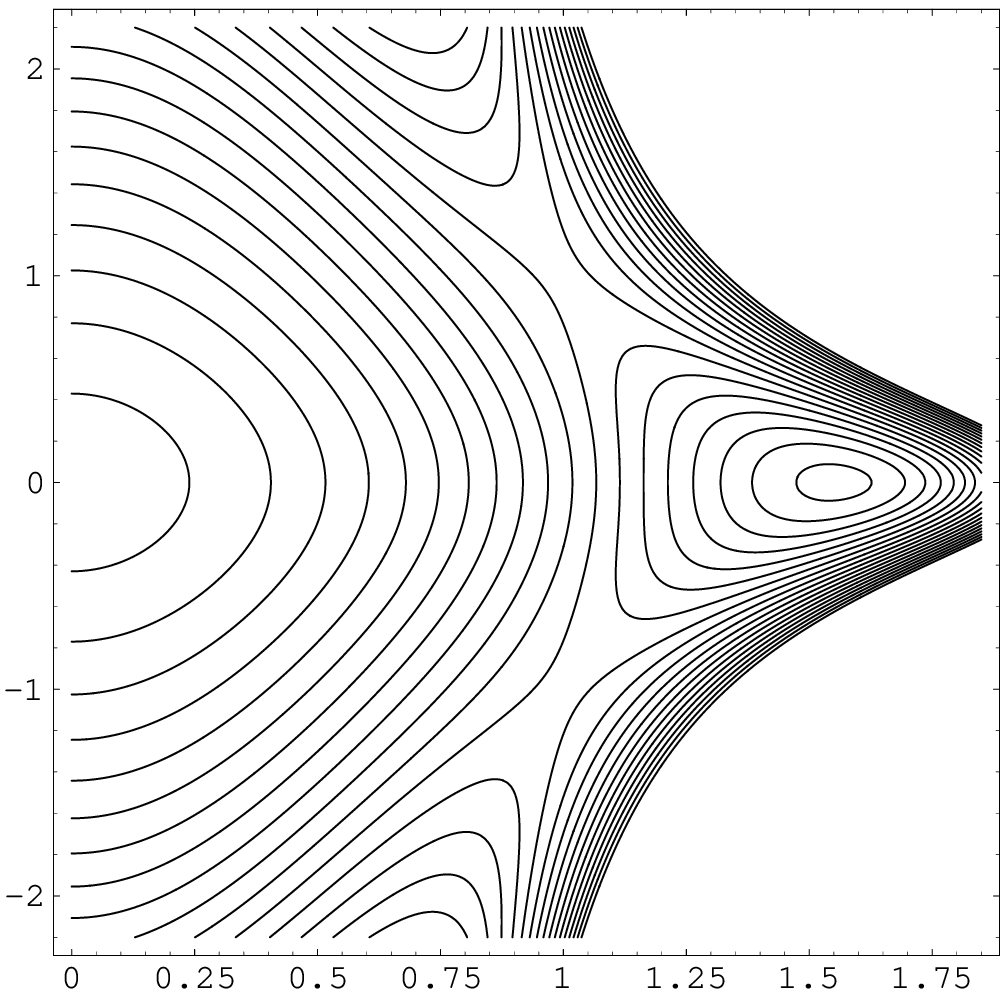}}
\medskip

\hspace*{5mm}
\begin{minipage}[htbp]{60mm}
{\small {\bf Figure 1:} 
Contour plot $V(x_1, x_2)$,  \\[-1mm]
slice: $x_1\equ y_1$, $x_2\equ y_2$; $x_2$ is vertical.}
\end{minipage}
\end{minipage}
\begin{minipage}[htbp]{80mm}
\resizebox{73mm}{!}{\includegraphics{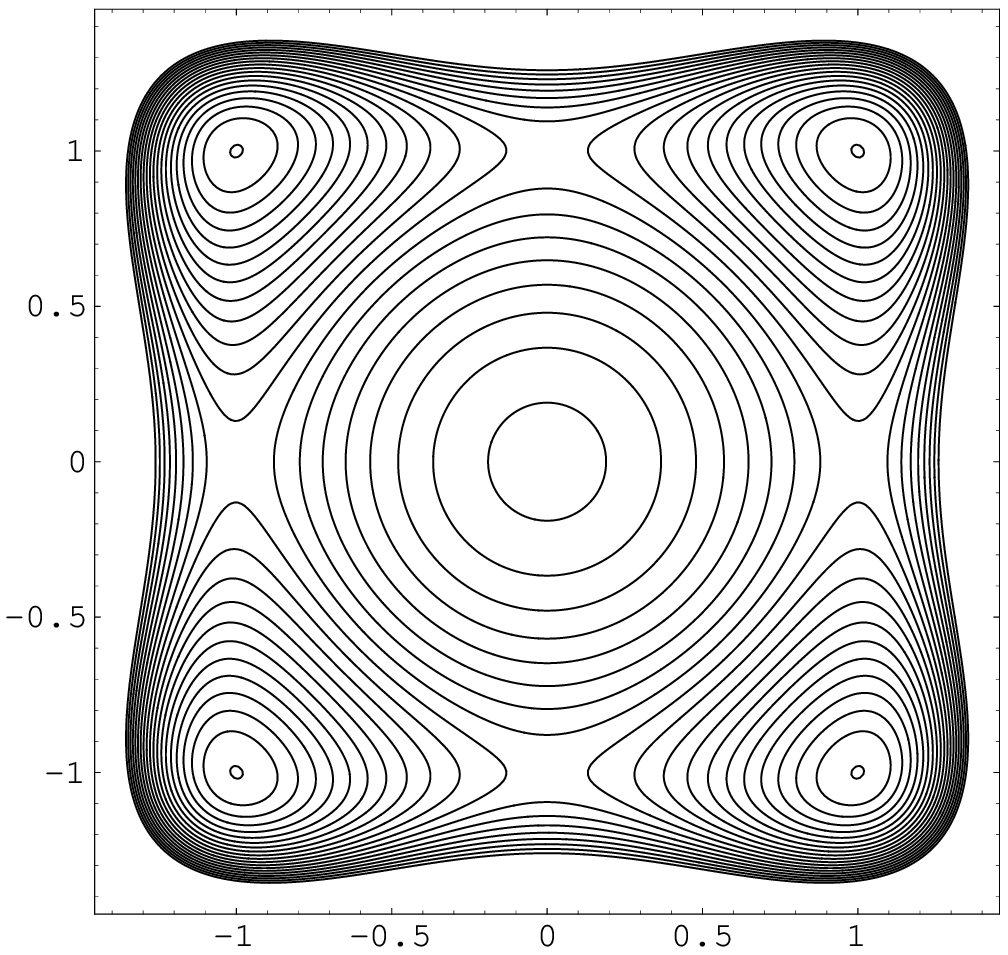}}
\medskip

\hspace*{8mm}
\begin{minipage}[htbp]{60mm}
{\small {\bf Figure 2:} 
Contour plot $V(x_1, y_1)$, \\[-1mm]
slice: $x_1\equ x_2$, $y_1\equ y_2$; $y_1$ is vertical. }
\end{minipage}
\end{minipage}

\addtocounter{figure}{2}

\bigskip
\bigskip

These figures exhibit the most interesting extrema in this truncation
of the theory. Details of these extrema are collected in
table~\ref{extrema}, in particular the remaining gauge- and
supersymmetry that they preserve. The unbroken supersymmetries are
encoded in the eigenvalues of the tensor $A_1$ from
\Ref{potential}, evaluated at the extremum \ci{GuRoWa86,NicSam01a}. The
number of preserved supersymmetries coincides with the eigenvalues
whose absolute value satisfies $|a_A|=1/(2g\adsr)$ with the AdS
radius $\adsr$ given by $\adsr=1/\sqrt{2V_{\Mat=\Mat_{\rm IR}}}$.  The
ratio of the central charge of the associated conformal field theory
and the CFT at the origin is given by \cite{HenSke98}
\be
\frac{c_{\rm IR}}{c_{\rm UV}} 
~=~ \sqrt{\frac{V_0}{V|^\vl_{\Mat=\Mat_{\rm IR}}} }
\;.
\la{ccVV}
\ee
All of these extrema are stable, i.e.\ upon linearizing the potential
around any of these extremal points, all $8n$ scalars satisfy the
Breitenlohner-Freedman bound $m^2 \adsr^2 \ge-1$. For the extrema
$a)$, $b)$, and $d)$, this is simply a consequence of the unbroken
supersymmetry; for the non-supersymmetric extremum $c)$, this can be
verified by explicit computation of the scalar fluctuations around the
extremum, see appendix~A.

\begin{table}[htbp]
\centering
\begin{tabular}{|l||c|c|c|c|} \hline
$~~(x_1, x_2, y_1, y_2)$ & 
{\footnotesize\begin{tabular}{c} \# multiplets\\
$n$\end{tabular}}
& 
{\footnotesize\begin{tabular}{c} unbroken\\gauge symmetry\end{tabular}}
&
{\footnotesize\begin{tabular}{c}unbroken\\supersymmetry \end{tabular}}
&
{\footnotesize\begin{tabular}{c}central charge\\
$c_{\rm IR}/c_{\rm UV}$ \end{tabular}}\\[0.1em]  
\hline\hline
$a)\;\; (0,0,0,0)$ & $0$ & $SO(4)\!\times\!SO(4)$ & $N=(4,4)$ 
& $1$  \\
\hline 
$b)\;\; (z_0,0,z_0,0)$ & $3$ & $--$ & $N=(1,1)$ 
& $\sqrt{2}-1$  \\
\hline 
$c)\;\; (1,1,0,0)$ & $4$ & $SO(4)$ & $--$ & $2/3$  \\
\hline 
$d)\;\; (1,1,1,1)$ & $4$ & $--$ & $N=(1,1)$ & $1/2$  \\
\hline 
\end{tabular}
\caption{\small Stable extrema with some remaining symmetry, 
$z_0=\sqrt{\sqrt{2}\!+\!1}$}
\label{extrema}
\end{table}

Although we have given these results for the particular value
$\Ga\equ1$ only, the qualitative features of the potential and its
extrema remain essentially unchanged for arbitrary~$\Ga$. Upon
decreasing the value of~$\Ga$, the shape of Figure~1 does not change
significantly, while Figure~2 is stretched along its vertical axis,
breaking the ${\mathbb Z}_4$ symmetry. In the limit $\Ga\ra0$, the
minima in the corners of Figure~2 (which correspond to the
supersymmetric extremum $d)$ in table~\ref{extrema}) disappear to
infinity. The top and bottom saddle points of Figure~2 also move off to
infinity, whereas the ones on left and right (which for $\Ga\equ1$
give the non-supersymmetric extremum $c)$ of table~\ref{extrema})
remain critical points of the potential, but become unstable below a
certain critical value of~$\Ga$. The other supersymmetric extremum
$b)$ behaves similarly under change of~$\Ga$.

\begin{figure}[htbp]
  \begin{center}
\epsfxsize=60mm
\epsfysize=50mm
\epsfbox{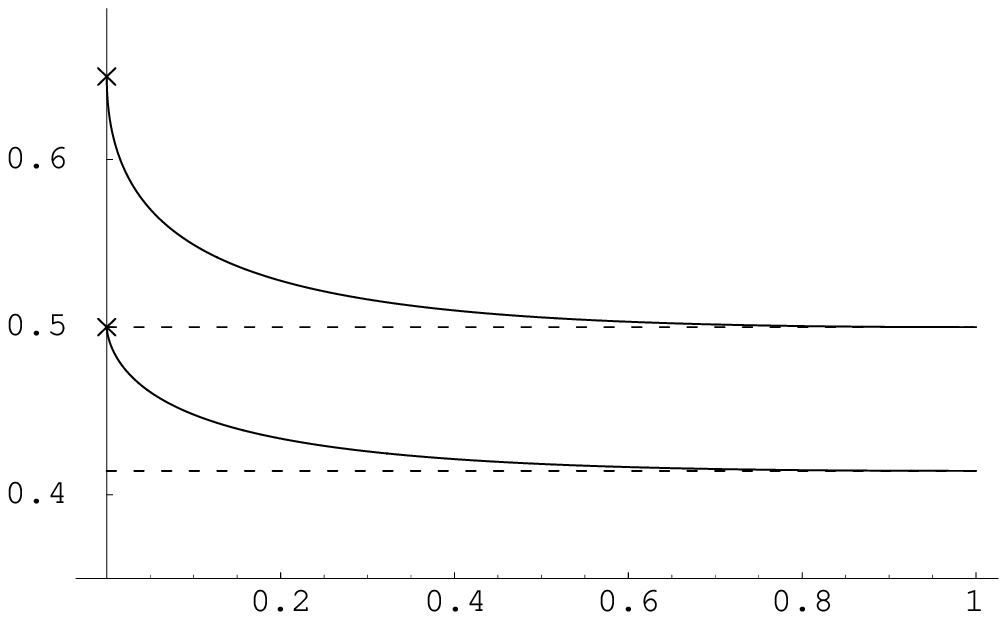}
  \caption{\small Ratios of central charges
                  as functions of $\alpha$, see (\ref{ccVV}).} 
 \label{cratio}
  \end{center}
\end{figure}

Summarizing, the two $N=(1,1)$ supersymmetric extrema in
table~\ref{extrema} have analogues for any value of $\Ga>0$, whereas
the precise sense of the limit $\Ga\ra0$ requires further
investigation.  The value of the potential at these extrema changes as
a function of $\Ga$, and so does its ratio to the value of the
potential at the origin \Ref{V00}.  We refrain from including the
somewhat lengthy explicit formulas here; as an illustration, the
ratios of the central charges of the associated IR boundary theories
to the central charge of the UV theory at the origin~\Ref{ccVV} are
plotted in Figure~\ref{cratio} for the two supersymmetric extrema as a
function of $\Ga$. At $\Ga\equ1$, these ratios give the values listed
in table~\ref{extrema}. In the limit $\Ga\ra0$ they tend to $1/2$ and
$3\sqrt{3}/8$, respectively. These values are never taken as both
extrema are absent in the theory obtained by naively setting
$\Ga\equ0$. The physical meaning of the limit $\Ga\ra0$ thus remains
to be fully understood. Recall that in the double D1-D5 system this
limit corresponds to sending the charges of one of the systems to
infinity.

\subsection{The supersymmetric extremum}

In the following we will mainly study the $N=(1,1)$ supersymmetric
extremum $d)$ from table~\ref{extrema}, which appears as a saddle
point in the slice of Figure~1 and as a minimum in the upper right
corner in Figure~2. The value of the potential at this point is
$V=-128\,g^2$, such that the AdS radius is given by $\adsr=1/(16g)$
(for definiteness we take $g>0$), and the ratio of central charges is
$c_{\rm IR}/c_{\rm UV}= 1/2\,$. As a first step we will compute the
spectrum of physical fields and their masses around this extremum. The
matrix $\Mat$ at this point takes the form
\be
\Mat_{\rm IR}~=~ \left(
\begin{array}{cccc}
\sqrt{3}\,I_4&0_{4\times(n-4)}&\sqrt{2}\,I_4&0_{4\times4}\\
0_{(n-4)\times4} & I_{n-4} & 0_{(n-4)\times4} & 0_{(n-4)\times4} \\
I_4&0_{4\times(n-4)}&\sqrt{\ft32}\,I_4&\sqrt{\ft12}\,I_4\\
-I_4&0_{4\times(n-4)}&-\sqrt{\ft32}\,I_4&\sqrt{\ft12}\,I_4
\end{array}
\right) \;,
\la{L0}
\ee
This preserves a larger subgroup than \Ref{Ginv}, namely
\ba
G_{\rm inv} &\equiv& SO(4)_{\rm inv}\times SO(n\mis4) 
\non 
&\subset&
\left(SO(4)^+ \times SO(4)^- \right) 
\times \left( SO(4)^{\vphantom +} \times SO(n\mis4) \right) 
\;,
\la{Ginv0}
\ea
where $SO(4)_{\rm inv}$ is embedded as the diagonal of the three
$SO(4)$ factors on the right hand side. Note that this is a {\em global}
remaining symmetry, i.e.\ the gauge group \Ref{gaugeG} is completely
broken at this extremum, and in agreement with the discussion above
there is no $R$-symmetry of the associated $N\equ(1,1)$ superconformal
field theory. Nevertheless, the spectrum may be organized under
\Ref{Ginv0}. With respect to $SO(4)_{\rm inv}$,
or rather $(SU(2) \times SU(2))_{\rm inv}$, the physical fields
decompose according to 
\ba
\mbox{gravitons}\,: && 2\cdot{\bf (1,1)} \non
\mbox{gravitini}\,: && 2\cdot{\bf (1,1)} + {\bf (3,1)} 
+ {\bf (1,3)} \non
\mbox{vectors} \,:&& 2\cdot\left( \vphantom{S^1} 
{\bf (3,1)} + {\bf (1,3)} \right) \non
\mbox{scalars} \,:&& 
2\cdot\left( \vphantom{S^1} 
{\bf (1,1)} + {\bf (3,1)} + {\bf (1,3)} + {\bf (3,3)}  
+  (n\mis4) \cdot {\bf (2,2)} \right)\non
\mbox{fermions} \,:&& 
2\cdot\left( \vphantom{S^1} 
{\bf (1,1)} + {\bf (3,1)} +  {\bf (1,3)} + {\bf (3,3)}  
+  (n\mis4) \cdot {\bf (2,2)}  \right) \;.
\la{Sinv}
\ea

The scalar masses around this extremum are obtained from computing the
fluctuations of the potential~\Ref{potential} around the
point~\Ref{L0}. As a result, we find that the two singlets in
\Ref{Sinv} acquire masses 
\be
m^2\,\adsr^2 = 
(\ft54, \ft{21}{4} ) \;.
\ee
The masses of the scalars in the ${\bf (3,3)}$ turn out to satisfy
$m^2\,\adsr^2 = -\ft34$, i.e.\ expressed in AdS units they coincide
with the masses around the central maximum, cf.~table~\ref{specASS}
and \Ref{lin34} --- however, the AdS units have of course changed with
respect to~$\Mat\equ I_{8n}$, the scale has shrunk to half: $L\equ
L_0/2$.  Half of the scalars in the ${\bf (2,2)}$ as well as those in
the ${\bf (3,1)}+{\bf (1,3)}$ representations become massless around
\Ref{L0}; the latter are the 12~Goldstone bosons associated with the
complete breaking of the gauge symmetry. The other half of the ${\bf
(2,2)}$ scalars turns out to saturate the Breitenlohner-Freedman
bound~$m^2\,\adsr^2 = -1$.

The vector fields satisfy first order (i.e.\ massive self-duality)
equations of motion due to their Chern-Simons coupling \Ref{LCS}, the
kinetic scalar term $\ft14 \sqrt{G} \, G^{\mu\nu}\,\CP_\mu^{Ir}
\CP_\nu^{\,Ir}$ serving as a mass term. Evaluating the latter at
\Ref{L0} and diagonalizing the resulting first order equations to
bring them to the form
\cite{ToPivN84}
\be
\Ge^{\mu\nu\rho}F_{\mu\nu} ~=~ 2m\sqrt{G} \, B^\rho \;,
\ee
we find that six of the vector fields come with masses $m\,\adsr =
\ft32$, and the other six with masses $m\,\adsr =\ft12$.

The calculation of the fermion masses finally requires the explicit
form of the fermionic part of the Lagrangian \Ref{L} which is given by
\cite{NicSam01b}
\ba
\CL_F&=& 
\ft12 \Ge^{\mu\nu\rho} \Bpsi{}^A_\mu D_\nu\psi^A_\rho 
-\ft12\,\I \sqrt{G}\, \Bchi^{\dA r}\Gg^\mu D_\mu\chi^{\dA r}
- \ft12\sqrt{G}\, \CP_\mu^{Ir}\Bchi^{\dA r} 
\GG^I_{A\dA}\Gg^\nu\Gg^\mu\psi^A_\nu
\la{LF}\\[1ex]
&&{}
+\ft12 g\sqrt{G} A_1^{AB} \,\Bpsi{}^A_\mu \Gg^{\mu\nu} \psi^B_\nu
+\I g\sqrt{G} A_2^{A\dA r} \,\Bchi^{\dA r} \Gg^{\mu} \psi^A_\mu
+\ft12 g\sqrt{G} A_3^{\dA r\, \dB s}\, \Bchi^{\dA r}\chi^{\dB s} 
\;.
\nn
\ea
The gravitini here are denoted by $\psi^A_\mu$, the matter fermions by
$\chi^{\dA r}$, indices $A, B, \dots$ and $\dA, \dB, \dots$ denote the
spinor and conjugate spinor representation, respectively, of 
the double cover of $SO(8)$. 
The scalar tensor $A_3^{\dA r\, \dB s}$ is defined
similar to $A_1$ and $A_2$ in \Ref{potential} above,
see~\cite{NicSam01b} for details. The gravitini masses in \Ref{LF} are
extracted from the eigenvalues of the tensor $A_1^{AB}$. Evaluating
$A_1$ at \Ref{L0} shows that two of the gravitini have masses
$m\adsr=\ft12$, corresponding to the two unbroken supersymmetries at
this extremum, while the other six become massive Rarita-Schwinger
fields with $m\adsr=1$.

The computation of the fermion masses is slightly more involved. They
are essentially encoded in the tensor $A_3^{\dA r\, \dB s}$; however
to properly take care of the super-Higgs effect and the mass-mixing
term in the Lagrangian, one must first fix the six broken
supersymmetry parameters by eliminating six of the fermion fields via
\be
\Gd\chi^{\dA r}~=~ A_2^{A\dA\,r} \,\Ge^A \;, 
\la{gaugeferm}
\ee
see \cite{FGPW99} for a complete discussion. Applied to our case, we
obtain fermion masses $m\adsr=\pm2$ for the two singlets in
\Ref{Sinv}, and masses $m\adsr=\pm1$ for one copy of the ${\bf
(1,3)}+{\bf (3,1)}$ while the other becomes massless due to the super
Higgs effect. The fermions in the ${\bf (3,3)}$ also become 
massless, whereas the ${\bf (2,2)}$ fermions come with masses
$m\adsr=\pm\ft12$.

We summarize the results of this section in table~\ref{specL0}. The
physical fields are organized under the global $SO(4)_{\rm inv}$
symmetry \Ref{Ginv0} and grouped into supermultiplets under the
$N=(1,1)$ superconformal symmetry on the boundary. The translation
between masses in three dimensions and conformal dimensions on the
boundary is given by
\be
\GD_{\pm} ~=~ 1 \pm \sqrt{1+m^2\adsr^2} \;,
\la{MDsc}
\ee
for the scalar fields \cite{Witt98}, and
\be
\GD  ~=~ 1 + |m|\adsr \;, 
\la{MDve}
\ee
for Rarita-Schwinger fields, matter fermions and vector fields with
Chern-Simons action \cite{HenSfe98,Volo98,KosRyt99,FGPW99}. As in
earlier work \cite{FGPW99}, we choose the sign in \Ref{MDsc} in
accordance with the group-theoretical structure.

\begin{table}[htbp]
\centering
\begin{tabular}{|c||l|c|c|} \hline
$SO(4)_{\rm inv}$ & fields & $m^2\adsr^2$ & $(h,\bar{h})$ \\
\hline\hline
{\bf (1,1)} & scalars & 
$\ft54, \ft{21}4$ & $(\ft74,\ft74)+(\ft54,\ft54)$ \\
\cline{2-4}
&fermions & $4$ & $(\ft74,\ft54)+(\ft54,\ft74)$ \\
\hline 
\hline 
{\bf (3,3)} & scalars & 
$-\ft34$ & $(\ft14,\ft14)+(\ft34,\ft34)$ \\
\cline{2-4}
&fermions & $0$ & $(\ft14,\ft34)+(\ft34,\ft14)$ \\
\hline 
\hline 
{\bf (1,3)} & fermions & $1$ & $(\ft34,\ft54)$ \\
\cline{2-4}
&vectors & $\ft14,\ft94$ & $(\ft14,\ft54)+(\ft34,\ft74)$ \\
\cline{2-4}
&massive gravitini & $1$ & $(\ft14,\ft74)$ \\
\hline 
\hline 
{\bf (3,1)} & fermions & $1$ & $(\ft54,\ft34)$ \\
\cline{2-4}
&vectors & $\ft14,\ft94$ & $(\ft54,\ft14)+(\ft74,\ft34)$ \\
\cline{2-4}
&massive gravitini & $1$ & $(\ft74,\ft14)$ \\
\hline 
\hline 
{\bf (2,2)} & scalars & 
$-1,0$ & $(1,1)+(\ft12,\ft12)$ \\
\cline{2-4}
&fermions & $\ft14$ & $(1,\ft12)+(\ft12,1)$ \\
\hline 
\end{tabular}
\caption{\small Spectrum around the supersymmetric extremum \Ref{L0}
in $N=(1,1)$ supermultiplets}
\label{specL0}
\end{table}

The super-Higgs effect occurs in the ${\bf (1,3)}$ and ${\bf (3,1)}$
where the 12 scalars and six of the matter fermions become massless
(and have not been included in table~\ref{specL0}), while the vectors
and the gravitino fields acquire mass. 

We emphasize that for this extremum not only the central charge but
also all conformal dimensions in the dual field theory, computed from
\Ref{MDsc}, \Ref{MDve}, come out to be rational. Of course, for a
finite number of primaries, rational charges and weights follow, but
one would not {\it a priori} expect there to only be a finite number
of primaries --- in fact, as can be clearly seen in the appendix, the
conformal weights are typically irrational also here in two
dimensions, just as in higher-dimensional examples.

\section{The kink solution}

We will now construct the kink solution of the gauged supergravity
\Ref{L} that interpolates between the central maximum $\Mat=I_{8n}$
and the supersymmetric extremum $\Mat=\Mat_{\rm IR}$~\Ref{L0}. As the
latter preserves one quarter of the supersymmetry, one is led to
search for a solution which preserves $N\equ(1,1)$ supersymmetry
throughout the flow.

\begin{figure}[htbp]
  \begin{center}
\epsfxsize=60mm
\epsfysize=40mm
\epsfbox{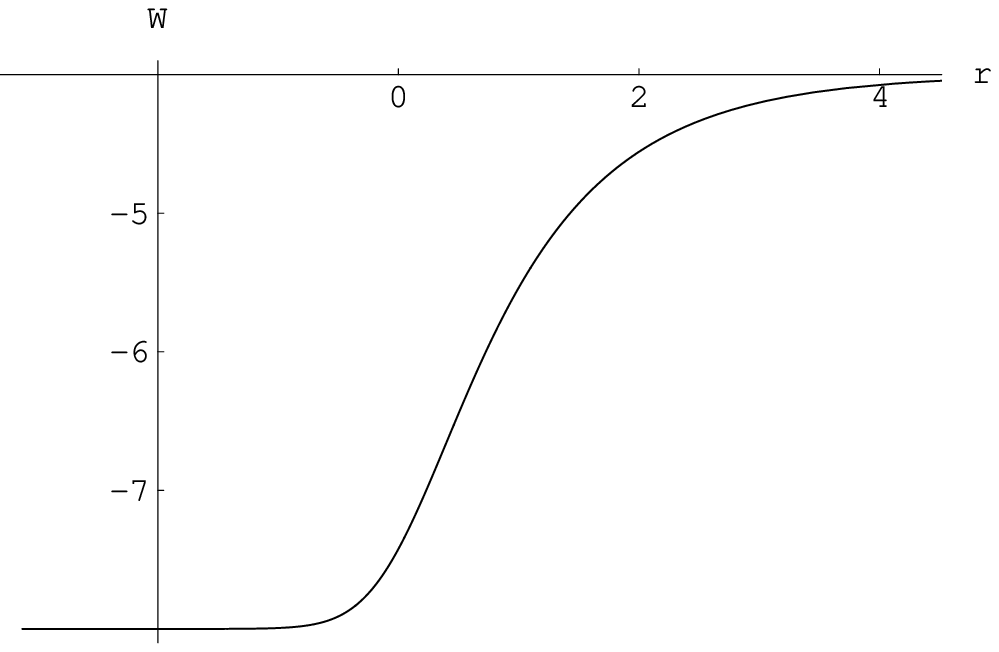}
  \caption{\small 
            The superpotential $W$ along the kink solution \Ref{kink}
(for $g\equ1/8$).} 
  \label{flowW}
  \end{center}
\end{figure}

For the metric we employ the standard domain wall ansatz
\be
ds^2~=~e^{2A(r)}\,\eta_{ij}\,dx^i dx^j - dr^2 \;,
\la{metric}
\ee
where $\eta_{ij}$ is the two-dimensional Minkowski metric. Pure AdS
geometry corresponds to linear behavior of $A(r)=r/\adsr\,$, with AdS
radius $\adsr$, cf.~\Ref{AdSrad}. For the scalar fields we use the
ansatz \Ref{matL} with $p_1\equ p_2\equ p_3\equ p_4\equ\pi/4$,
$q_1\equ q_2\equ q_3\equ q_4\equ q(r)$. This corresponds to switching
on one of the two singlet scalars from \Ref{Sinv}.  A scalar that
acquires a radial dependence in the flow is usually referred to as an
``active'' scalar, whereas the scalars that are zero in the background
are ``inert''.  In this truncation, the scalar potential \Ref{Vex}
reduces to a potential for just the active scalar $q$:
\be
V_0=-\frac{g^2}{16}\,
\Big((3 + \cosh 2q)^2  (21 + 12 \cosh 2q - \cosh 4q) \Big)
\;,
\la{VQ}
\ee
while the kinetic term is just $\sqrt{G}\,G^{\mu\nu}\,\dd_\mu q
\dd_\nu q $\,. Straightforward calculation shows that the tensor $A_1$
from \Ref{potential} can be diagonalized with $q$-independent
eigenvectors, to take the form
\ba
A_1 &=& \diag( -X, -X, -X, X, X, X, -W, W ) \;,
\non[1ex]
&& X=(1\pls\cosh^2q)^2 \;,\non
&& W=-\frac{1}{8}\, (13 + 20 \cosh 2q - \cosh 4q) \;.
\la{W}
\ea
As usual, $W$ represents the superpotential of the scalar potential
$V_0$, which may be derived from the former as  
\be
V_0~=~ \frac14\,g^2\, 
\left(\frac{\dd W}{\dd q}\right)^2 - 2g^2\,W^2
\;.   
\ee
We denote the corresponding eigenvectors of $A_1$ by $v_\pm$:
\be
A_1 \, v_\pm ~=~ \pm W \, v_\pm \;.
\ee
The supersymmetric critical points $\Mat=I_{8n}$ and $\Mat=\Mat_{\rm
IR}$~\Ref{L0} of the scalar potential $V_0$ are also critical points
of the superpotential $W$. In particular, this ensures that these
points are non-perturbatively stable~\cite{Town84,SkeTow99}. The
supersymmetric kink solution is now derived by solving the Killing
spinor equations for the gravitino and the matter
fermions~\cite{NicSam01b}
\ba
\Gd_\Geps \psi^A_\mu &=& 
D_\mu \Geps^A + \I g A_1^{AB}\Gg_\mu\,\Geps^B  ~\stackrel{!}{\equiv}~ 
0 \;, \non
\Gd_\Geps \chi^{\dA r} &=& \left(
\ft12\I\GG^I_{A\dA}\Gg^\mu \CP_\mu^{Ir} +g A_2^{A\dA r} \right)
\Geps^A  ~\stackrel{!}{\equiv}~ 0 \;.
\la{KS}
\ea
With the following ansatz for the Killing spinor $\Geps$
\be
\Geps ~=~ 
F_+(r)\,v_+\, (1\mis\I\Gg^r)\, \eta_0  
+ F_-(r)\, v_-\, (1\pls\I\Gg^r)\, \eta_0 \;,
\la{Kspinor}
\ee
where $\eta_0$ denotes a constant real 3d spinor, the
second equation from \Ref{KS} reduces to
\be
\frac{d q}{d r} ~=~ \frac{g}{2}\,\frac{d W}{d q} ~=~
-\frac{g}{4}\, ( 10 \sinh2q - \sinh4q ) \;.
\la{QW}
\ee
This equation may be {\em analytically} solved as the root of a cubic
equation:
\be
\frac{(5-y)(y+1)^2}{(y-1)^3} ~=~ c_1\, e^{24gr} \;,
\qquad
\mbox{where} \quad
y=\cosh2q \;,
\la{kink}
\ee
and with integration constant $c_1$ which may be absorbed into a shift
of $r$ and will be set to $c_1\equ2$. This flow runs from the central
maximum ($y\equ1$) at $r=\infty$ to the supersymmetric extremum
\Ref{L0} ($y\equ5$) at $r=-\infty$. The behavior of the superpotential
$W$ along this flow is depicted in figure
\ref{flowW}. Following~\cite{GPPZ98,FGPW99,AGPZ00}, the holographic
$C$ function is defined to be proportional to $-1/W$. We emphasize
that in our two-dimensional setting, upon computing 2-point
correlation functions for the energy-momentum tensor one may
eventually compare this holographic definition to Zamolodchikov's
definition of the $C$ function~\cite{Zamo86}.

It remains to solve the first equation in the Killing spinor
equations~\Ref{KS}. Substituting \Ref{Kspinor}, we obtain
\ben
\dd^{\vphantom b}_\mu \Geps^A -
\ft14\,\I\,\Go_\mu{}^{ab}\,\Ge_{abc}\,\Gg^c\, \Geps^A  
+ \I g W\left( F_+(r)\,v_+^A\, (1\mis\I\Gg^r) 
- F_-(r)\, v_-^A\, (1\pls\I\Gg^r) \right) \eta_0
{}~\stackrel{!}{\equiv}~ 0 \;,
\een
where the spin connection $\Go_\mu{}^{ab}$
introduces the metric into the equation.
This yields
\be
\frac{d A}{d r}~=~-2g\,W \;,\qquad
F_\pm ~=~ c_\pm\,e^{-A/2} \;,
\la{AW}
\ee
with constants $c_\pm$. Using \Ref{QW}, \Ref{kink}, this equation 
may explicitly be integrated as a function of the scalar $q(r)$:
\be
e^{6A(r)} ~=~ \frac{(5-y)^{4}}{2\,(y+1)(y-1)^6} \;,
\la{Ar}
\ee
where the integration constant has been fixed from asymptotics at
$r\ra\infty$. Asymptotically, $A(r)$ goes as $8gr$ for $r\ra\infty$
and as $16gr$ for $r\ra-\infty$, in accordance with the pure AdS
behavior of \Ref{metric}. That is, setting
$g=1/8$ yields pure AdS with length scale $L_0=1$ 
at $r\ra\infty$, which 
is what one expects from equation (\ref{AdSrad}).

To summarize, we have found an analytical solution for the domain wall
spacetime given by \Ref{kink}, \Ref{Ar}, admitting two Killing spinors
of the form \Ref{Kspinor}, which interpolates between the central
maximum and the extremum \Ref{L0}. Previously constructed holographic
flows between conformal endpoints in higher dimensions could only be
given numerically~\cite{GPPZ98,DisZam99,FGPW99}, preventing the
computation of correlation functions in the boundary theories.

To analyze the near-boundary asymptotics of \Ref{kink}, it is
convenient to introduce the new variable $\rho$~\cite{HenSke98}
\be
\rho= e^{-2r/\adsr_0} = e^{-16gr} \;,
\ee
such that the line element \Ref{metric} becomes
\be
ds^2~=~
\frac{1}{\rho}\,e^{2(A(r)-r/\adsr_0)}\,\eta_{ij}\,dx^i dx^j 
- \frac{\adsr_0^2 }{4\rho^2}\, d\rho^2
\;.
\ee
The asymptotics of the kink solution \Ref{kink}, \Ref{Ar} close to
$\rho=0$ (i.e.\ close to the central maximum
of the scalar potential) is then given by
\ba
q(\rho) &=&
\rho^{\frac14} 
\left( 
1
+ \frac1{12}\,\sqrt{\rho}
- \frac{13}{160}\, \rho  + \CO( \rho^{3/2} ) \right)
\;,
\non
e^{2A(\rho)} &=& 
\frac{1}{\rho} 
\left(1
- 2\, \sqrt{\rho} 
+ \frac{7}{4}\,\rho + \CO( \rho^{3/2} ) \right)
\;,
\non
W &=& -4 \left( 1 + q^2 +\frac{1}{12}\,q^4 + \CO( q^6 ) \right)
\;.
\la{asym}
\ea
This behavior of the active scalar field $q(\rho)$ shows that as
anticipated in the introduction, this solution --- interpreted as a
holographic RG flow --- corresponds to a deformation of the UV
conformal field theory by a relevant operator of dimension
$\GD=\ft32$, rather than to a vev~\cite{KleWit99}.  Indeed, the lowest
order power-law behavior of $q(\rho)$ is that expected from standard
arguments: $q(\rho) \sim \rho^{(d-\Delta)/2}=\rho^{1/4}$ (cf.\
equation (2.10) in \cite{BiFrSk01}).  It may be worth pointing out
that the appearance of noninteger powers of $\rho$ in the parentheses
above is in general limited to half-integer powers; the expansion is
ultimately an expansion in the original variable $r$, which for an AdS
radius of $L_0\equ1$ is just $r\equ-\log \sqrt{\rho}$. Also we warn
the reader that a fair number of expressions in the literature
actually degenerate for $d=2$, the case treated here.

As mentioned in the introduction, the tentative conjecture is that
this active scalar is dual to a mass operator for chiral superfields
in the large $N=4$ boundary theory.  Let us now make this a little
more precise: consider the decoupled worldvolume theory on one
D-brane.  The spacetime theory on $N$ unordered branes is then the
symmetric product of $N$ such theories, i.e.\ orbifolded by the
permutation among the branes.  For definiteness, take the
two-dimensional large $N=4$ theory with $c=\tilde{c}=3$ constructed
from one scalar field and four fermions on each chiral side.  In $N=2$
language, and bosonizing one pair of fermions on each side, these
fields fill out two hypermultiplets.  (Giving vevs to these
hypermultiplets would describe motion on the Higgs branch
\cite{Witt97, DiaSei97}, however as pointed out above our flow is not
a vev flow).  This bosonizing hides half of the $SU(2)\times SU(2)$
$R$-symmetry, leaving the diagonal $SU(2)$ rotating within the
hypermultiplets, and the $SU(2)$ rotating the hypermultiplets into
each other.  Adding a term which is bilinear in chiral superfields and
a singlet under the combined action of these two symmetries, yields a
mass term that breaks supersymmetry to $N=(1,1)$ and the effective
theory when the massive fields are integrated out has
$c=\tilde{c}=3/2$.  Such a mass term then appears to be the operator
that couples to the active scalar field in our flow.  The symmetries
of this term are consistent with the representation content; our
active scalar is a singlet under the remnant symmetry of
simultaneously rotating both three-spheres accompanied by a rotation
of the matter multiplets.  The anomalous dimension $\Delta=3/2$ cannot
be explained by an argument of this type, but one way the conjecture
could be checked is if one could find an exact beta function
\cite{NSVZ83} due to the broken conformal symmetry.

A difficult issue in this correspondence is to distinguish, in
supergravity, between the bound that short multiplets saturate in
representations of near-boundary bulk isometries and the bound in the
boundary CFT
\be
kh \geq (\ell_L^+ - \ell_L^-)^2 + k^-\ell_L^+ + k^+ \ell_L^- + u^2 \;,
\label{hlu}
\ee
where $k=k^+ + k^-$ is the sum of the levels of the two $SU(2)$ factors,
and $u$ is the $U(1)$ charge. For large $k$, the nonlinearities
disappear from view: setting $\alpha=k^+/k^-$ and considering
low-lying $\ell^+$, $\ell^-$ one recovers (\ref{hl}).  Since the
circle in the $S^3 \times S^3\times S^1$ compactification is taken
small in the large $N$ limit, it is difficult to see how one would
distinguish states of different $U(1)$ charge in supergravity.

\section{Counterterms and one-point functions} 

In this section, we compute counterterms for scalars and 1-point
functions of their CFT duals
following~\cite{dHSoSk01,BiFrSk01,BiFrSk01a}. Together with the
fluctuation xequations derived in the next section, this in principle
yields 2-point functions of the CFT operators throughout the flow.
Although correct correlation functions for many operators were
obtained long before the aforementioned papers, more difficult cases
remained fraught with problems (see e.g.~\cite{DWoFre00}). Progress
was reported in \cite{ArFrTh00}, followed by the
emergence of a coherent picture
in~\cite{dHSoSk01,BiFrSk01,BiFrSk01a}. In the current point
of view, there are two main ideas to keep in mind; first, the old
realization that counterterms are to be introduced on a regulating
surface close to the
boundary~\cite{BalKra99,HenSke98,Myer99,dHSoSk01}.  Second, the
addition of finite covariant counterterms, corresponding to a
renormalization scheme that preserves
supersymmetry~\cite{BiFrSk01,BiFrSk01a}.  In keeping such a scheme the
use of covariant counterterms becomes especially crucial since the
latter differ from the non-covariant counterterms in their finite
parts.  In this paper, we need not worry about finite counterterms
since we only discuss uncoupled fluctuations here (inert scalars), but
they will become important in correlation functions involving the
active scalar~\cite{BerSam02}.

The AdS/CFT correspondence, extended to asymptotically AdS space,
posits that CFT correlation functions 
in the large $N$, strong coupling limit can be computed
from~\cite{GuKlPo98,Witt98}
\be
\langle e^{\int \hat{\phi} \sO} \rangle_{\hat{g}}
= e^{-S_{\rm sugra}[\phi,g]} \;,
\ee
where $S_{\rm sugra}$ is the classical supergravity action, and the
hatted quantities are the scaled Dirichlet data, i.e.\ where the
(often divergent) dependence on the AdS radius is factored out
\cite{dBVeVe00}.  (If we would not have taken the supergravity limit,
the right hand side would be the full string partition function.)  We
now proceed to perform the near-boundary analysis (i.e.\ around
$\rho\equ0$) for the inert scalars in the kink spacetime. For the
order we are interested in, it is most convenient to first linearize
their equations of motion.

According to~\Ref{Sinv}, under $SO(4)_{\rm inv}$ there is one inert
scalar in the ${\bf (1,1)}$, two copies of the ${\bf (3,3)}$, and two
times $(n\mis4)$ copies of the ${\bf (2,2)}$. We shall denote these
sectors by $\{ {\bf 1}, {\bf 9{\rm (1)}}, {\bf 9{\rm (2)}}, {\bf 4{\rm
(1)}}, {\bf 4{\rm (2)}} \}$, respectively. We may accordingly express
the matrix~$\Mat$ as
\be
\Mat=\Mat_q\,\Mat_\phi\;,\qquad
\mbox{with}\quad \Mat_\phi = 
\exp \sum_\Si \phi^\Si\, Y^\Si  \;,
\la{infsc}
\ee
where $\Mat_q$ is given by $\Mat_0$ from \Ref{matL} with $p_i=\pi/4$
and $q_i=q(r)$ while the sum in $\Mat_\phi$ is a short hand notation
for the scalars in the different representations noted above, i.e.\
the index $\Si$ runs over $\{{\bf 1}, {\bf 9{\rm (1)}}, {\bf 9{\rm
(2)}}, {\bf 4{\rm (1)}}, {\bf 4{\rm (2)}} \}$ and the $Y^\Si$ denote
the corresponding linear combinations of noncompact generators
$Y^{Ir}$. With this ansatz, the current~\Ref{current} becomes
\be
\Mat^{-1}\dd_\mu\Mat ~=~
\Mat^{-1}_\phi\dd_\mu\Mat^\vl_\phi + 
\Mat^{-1}_\phi(\Mat^{-1}_q\dd_\mu\Mat^\vl_q)\,\Mat^\vl_\phi \;.
\la{currpq}
\ee
Expanding the kinetic term in Lagrangian \Ref{L} to second order in
the inert scalar fluctuations $\phi^\Si$, thus gives a $q$-independent
kinetic term for the $\phi^\Si$, whereas the second term in
\Ref{currpq} contributes to the potential for the $\phi^\Si$ upon
inserting~\Ref{QW} for $\Mat^{-1}_q\dd_\mu\Mat^\vl_q$. Substituting
\Ref{infsc} and \Ref{currpq} into the Lagrangian \Ref{L}, leads after
some computation to
\be
\CL ~=~  \CL_q + \CL_{\phi} \;.
\la{Lqua}
\ee
The different parts of the Lagrangian are given by
\ben
\CL_q ~=~ 
\sqrt{G} G^{\mu\nu} \,\dd_\mu q \,\dd_\nu q -g^2\sqrt{G} \,V_0(q) \;,
\een
describing the active scalar $q$, with the potential $V_0$ from
\Ref{VQ} above, while the fluctuations of the inert scalars are
described by 
\be
\CL_{\phi} ~=~ 
\ft14\sqrt{G}\,{\textstyle \sum_\Si} \,
G^{\mu\nu}\,\dd_\mu\phi^\Si\,\dd_\nu\phi^\Si
- g^2\sqrt{G}\,{\textstyle \sum_{\Si,\Sj}} \,
V_{\Si\Sj}(q)\,\phi^\Si\,\phi^\Sj \;.
\la{Lin}
\ee
Miraculously, the potential $V_{\Si\Sj}(q)$ may be diagonalized with
$q$-independent eigenvectors in each of the two-fold degenerate
representation sectors ${\bf 4}$ and ${\bf 9}$, respectively, such
that their equations of motion decouple.~\footnote{A similar miracle
in~\cite{BWFP01} was uncovered by appealing to supersymmetry.}
Specifically, we find $V_{\Si\Sj}(q) = \delta _{\Si\Sj} V_{\Si}(q) $
with
\ba
V_{\bf 1} &=& \ft1{16}\,
( -45 - 160\,y + 10\,y^2 + 3\,y^4 )
\non
V_{\bf 9{\rm (1)}} &=& - \ft1{4}\,
( 17 + 30\,y + y^2 )
\non
V_{\bf 9{\rm (2)}} &=& \ft1{16}\,
(y+1)( -93 + 13\,y - 19\,y^2 + 3\,y^3) 
\non
V_{\bf 4{\rm (1)}} &=& \ft1{16}\,
(y+1)(y-5)(17 + 4\,y + 3\,y^2)
\non
V_{\bf 4{\rm (2)}} &=& -\ft1{4}\,
(3+y)(7+5y) \;,
\la{Vi}
\ea
where $y\equ\cosh2q$, cf.~\Ref{kink}.  The equations of motion for the
inert scalar fluctuations implied by \Ref{Lqua} thus decouple to
\be
\dd_\mu(\sqrt{G}G^{\mu\nu}\dd_\nu\phi^\Si) ~=~ 
-4g^2\sqrt{G}\, V_\Si(q)\, \phi^\Si \;.
\la{eomL}
\ee

We need to expand this equation and the Einstein field equations
around $\rho\equ0$. As seen in the previous section~\Ref{asym}, square
roots of the radial variable $\rho$ appear in the background; this
will also be the case for fluctuations of our scaled Dirichlet data.
Square roots are to be expected for a $\Delta=3/2$ flow in two
dimensions, since in general back reaction appears at order $d-\Delta$
for fields dual to relevant operators~\cite{dHSoSk01}. To regulate the
divergence of the action at $\rho=0$, we follow the standard
prescription of cutting off the bulk integral at $\rho=\epsilon$ and
including boundary terms at this radius.  Our metric and scalar
ans\"atze read
\ba
G_{ij}(x,\rho) &=& \rho^{-1} g_{ij}(x,\rho) \;,\non
g_{ij}(x,\rho) &=& g_{(0)ij}(x) + 
\sqrt{\rho}\; g_{(1)ij}(x) + 
\rho \, g_{(2)ij}(x)  +  \rho \log \rho \, h_{(2)ij}(x) +
\sO(\rho^{3/2}) \;,\non
\phi(x,\rho) &=& 
\rho^{1/4}\varphi(x,\rho) \;,\non
\varphi(x,\rho) &=& 
\varphi_{(0)}(x) + \sqrt{\rho}\, \varphi_{(1)}(x)+
\sqrt{\rho} \log \rho \, 
\psi_{(1)}(x)+ \rho\, \varphi_{(2)}(x)  + \sO(\rho^{3/2}) \;.
\ea
The logarithmic terms are needed at the given orders, and only there,
because the terms at those orders in the naive ans\"atze are not
determined by the equations of motion.  The active scalar in the
background is
\bean
q &=& \rho^{1/4}\acq =\rho^{1/4} (\acq_{(0)}+
\sqrt{\rho}\, \acq_{(1)}+ \rho \, \acq_{(2)} + \sO(\rho^{3/2}) )\;.
\eean
We relabel
\be
q \rightarrow \ft1{\sqrt{2}}q \;, \quad
\phi \rightarrow  \sqrt{2}\phi \;,\quad
V_0, V_\Si \rightarrow -V_0, -V_\Si \;, \quad
g_{ij} \rightarrow \; L_0^2 g_{ij} \; ,
\ee
and perform a Wick rotation to obtain a canonical Lagrangian
\[
\sqrt{G}^{-1} \CL = 
\ft1{2\kappa^2} R +
\ft12 G^{\mu \nu} \partial_{\mu} q \partial_{\nu} q 
+ V^q+
\ft12 G^{\mu \nu} \partial_{\mu} \phi \partial_{\nu} \phi  + V^{\phi}
\;,
\]
where we have restored the gravitational coupling $\kappa$. 
Notice that AdS with
length scale 1 is recovered by $g^2\kappa^2=1/32$. We consider
Einstein's equations with the stress-energy tensor
\[
T_{\mu \nu} = \partial_{\mu}\phi  \partial_{\nu} \phi
-G_{\mu \nu} (\ft12 (\partial \phi)^2 + V^{\phi}) \; ,
\]
and the same for $q$. The  order $\rho^{-1/2}$ terms yield
\ba
\tr (g_{(0)}^{-1}g_{(1)}) &=& 
 -{\kappa^2  }(\acq_{(0)}^2
+ \varphi_{(0)}^2) \;,
\ea
while the $\rho^{-1/2}$ term from \Ref{eomL} gives
\ba
\psi_{(1)} &=& -\ft18 (\tr g_{(0)}^{-1} g_{(1)})\varphi_{(0)} +
\ft14\acq_{(0)}^2\varphi_{(0)} \;.
\ea
Setting $\varphi=0$, we can check that our background~\Ref{asym}
indeed satisfies these equations.  The scalar curvature $R$ does not
appear here; it enters at order $\rho^0$. We can now write down the
regularized on-shell action and apply 
the fixed-background formalism.
\ba
S_{M,{\rm reg}}
&=& -\int_{\rho=\epsilon}
\!d^2 x \sqrt{g} \; \epsilon^{-1/2}(\ft14 \varphi^2 + 
\epsilon \varphi \dd_\epsilon\varphi) \non
&=& \int_{\rho=\epsilon}
\!d^2 x \sqrt{g_{(0)}}
\; (\epsilon^{-1/2} a^{\rm M}_{(0)} -\log \epsilon \;
a^{\rm M}_{(1)}
+ \sO(1))\;,
\ea
where
\[
a^{\rm M}_{(0)} = -\ft14 \varphi_{(0)}^2 \;  , \qquad
a^{\rm M}_{(1)} = \psi_{(1)} \varphi_{(0)}
\;.
\]
To take all quantities back to the surface at $\rho=\epsilon$, the
prescription is to perturbatively invert the relations between
$\varphi_{(0)}$ and $\varphi$, and $g_{(0)}$ and $g$. For us, the
change from $\sqrt{g_{(0)}}$ to $\sqrt{g}$ only contributes to finite
terms.  We then have the renormalized action, with the induced metric
$\sqrt{\gamma} = \epsilon^{-1} \sqrt{g}$ :
\ben
S_{{\rm M,ren}} = \lim_{\epsilon \rightarrow 0}
\left (S_{\rm bulk}(\rho \leq \epsilon)  + 
 \int_{\rho=\epsilon}
\! d^2 x \sqrt{\gamma} \left( \ft14\,\phi^2(x,\epsilon)+
\frac{2+\kappa^2}{16}\,
{\log \epsilon \over \epsilon} \; q^2 \phi^2 (x,\epsilon)\right)
\right)
\;.
\een
Computing the 1-point function is now simple; there is a contribution
both from the regularized action and from the quadratic counterterm
\be
\langle \sO_\phi \rangle = 
\lim_{\epsilon \rightarrow 0}
{ 1\over \sqrt{g_{(0)}}} {\delta S \over \delta\varphi_{(0)}}
= \lim_{\epsilon \rightarrow 0}
\left({1 \over \epsilon^{3/4}}
{ 1\over \sqrt{\gamma} } {\delta S \over \delta\phi }
\right)
= -2(\varphi_{(1)}+ \psi_{(1)})\;,
\ee
where we emphasize that we are neglecting finite counterterms.  In the
background, the inert scalar $\phi$ is of course zero, so this vev
does not constitute a useful check on the computations.  A nontrivial
check is afforded by the similar computation for the active scalar,
including finite counterterms~\cite{BerSam02}.

\section{Fluctuation equations}

The near-boundary analysis of the previous section is not sufficient
to calculate 2-point correlation functions of the associated operators
in the boundary theory. In this section we will compute the quadratic
fluctuations of the full Lagrangian~\Ref{L} around our flow
solution~\Ref{kink}. The 2-point correlation functions may then be
extracted from the properly normalized solutions to these
equations~\cite{DWoFre00,BWFP01,BiFrSk01,Muck01,BiFrSk01a}. As in
those cases it turns out that the analysis is most conveniently
performed in new so-called horospherical coordinates in which the line
element~\Ref{metric} takes the form
\be
ds^2~=~e^{2A(z)}\,(\eta_{ij}\,dx^i dx^j - dz^2) \;,
\qquad\quad\mbox{i.e.}\quad
\frac{dz}{dr}=e^{-A} \;.
\la{conmet}
\ee

For simplicity, we restrict to the fluctuations of inert scalars
and vector fields, postponing the active scalar and metric
fluctuations to future work. The analysis is greatly simplified by the
remaining $SO(4)_{\rm inv}$ symmetry~\Ref{Ginv0} which organizes the
spectrum. At this linearized level, the scalar fields split into the
gauge invariant sectors ${\bf (1,1)}$, ${\bf (3,3)}$, and ${\bf
(2,2)}$, whereas the scalars in the ${\bf (3,1)}+{\bf (1,3)}$ couple
to the vector fields and are shifted under the action of the gauge
group~\Ref{gaugeG}. We treat the two cases separately.

\subsection{Inert scalars}

We have shown in the previous section, that the equations of motion
for the inert scalar fluctuations decouple into second order
differential equations~\Ref{eomL} with potentials~\Ref{Vi}. With the
ansatz $\phi^\Si = e^{-i (p\cdot x)} \, e^{-A(z)/2} \, R^\Si(z) $, the
Laplace equations \Ref{eomL} turn into
\be
(-\dd_z^2 + \CV_\Si\,) \,R^\Si = p^2 R^\Si \;,
\la{eomR}
\ee
with the coordinate $z$ from \Ref{conmet} and an effective potential 
\be
\CV_\Si ~=~ \ft12 A''(z) + \ft14 A'(z)^2 + 4g^2 e^{2A}\,V_\Si 
{}~=~
e^{2A} g^2 \left( 3 W(q)^2 - \ft12 W'(q)^2 + 4\,V_\Si\right) 
\;.
\la{Veff}
\ee
With \Ref{W}, \Ref{Vi} we obtain the following effective potentials:
\ba
\CV_{\bf 1} &=& 
\ft1{16}\,g^2 e^{2A}\,(y-1)^2\,( 167 + 34\,y + 7\,y^2 )
\non
\CV_{\bf 9{\rm (1)}} &=& 
-\ft5{16}\,g^2 e^{2A}\,(y-1)^2\,(y-5)(y+3)
\non
\CV_{\bf 9{\rm (2)}} &=& 
\ft1{16}\,g^2 e^{2A}\,(y-1)^2\,(y-5)(7y+5) 
\non
\CV_{\bf 4{\rm (1)}} &=& 
\ft1{16}\,g^2 e^{2A}\,(y-1)^2\,(7 + 2\,y + 7\,y^2 )
\non
\CV_{\bf 4{\rm (2)}} &=& 
-\ft1{16}\,g^2 e^{2A}\,(y-1)^2\,( -11 - 10\,y + 5\,y^2 ) \; .
\la{Veffs}
\ea
Note that according to \Ref{Ar}, the factor $e^{2A}$ diverges as
$(y\mis1)^{-2}$ near $y\equ1$, such that all the effective potentials
tend to a finite value at the UV boundary. At the other end of the
flow ($y\equ5$) the effective potentials vanish. Their behavior along
the flow is depicted in Figure~\ref{FVeff}.

\begin{figure}[htbp]
\begin{center}
\epsfxsize=45mm
\epsfysize=35mm
\epsfbox{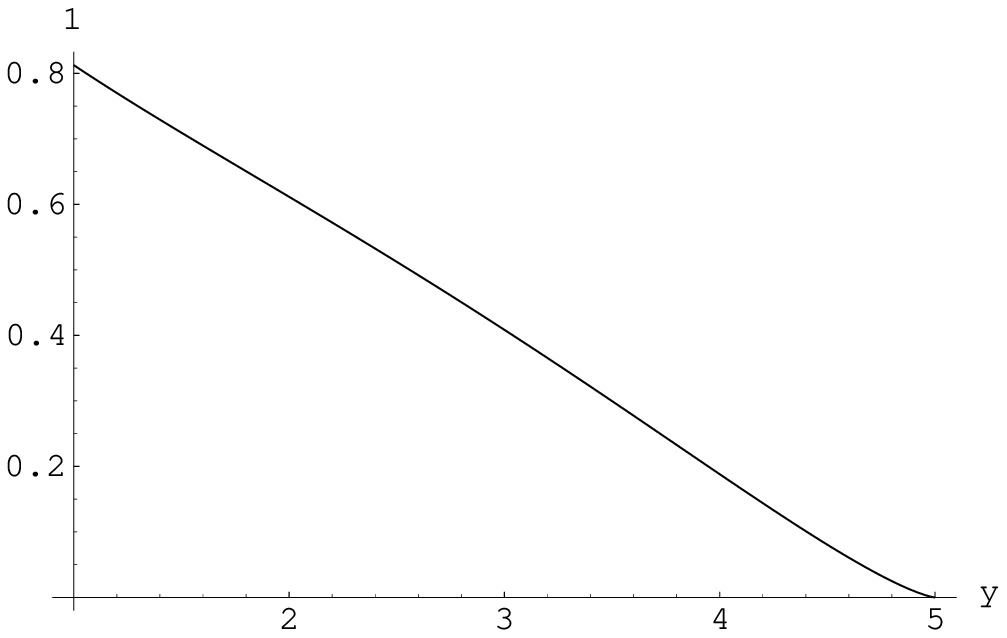}
\qquad
\epsfxsize=45mm
\epsfysize=35mm
\epsfbox{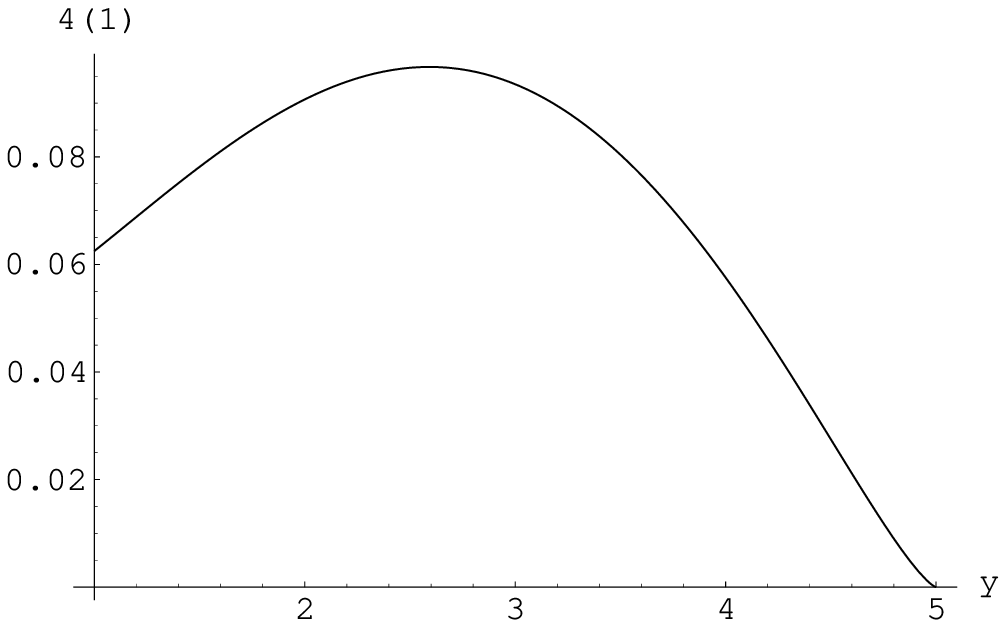}
\qquad
\epsfxsize=45mm
\epsfysize=35mm
\epsfbox{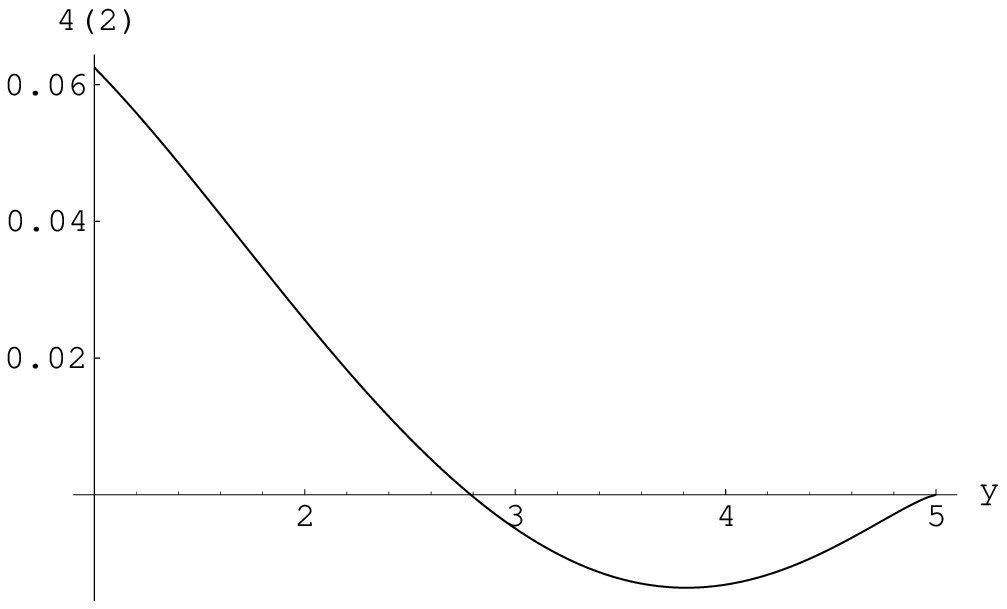}
\end{center}
\begin{center}
\epsfxsize=45mm
\epsfysize=35mm
\epsfbox{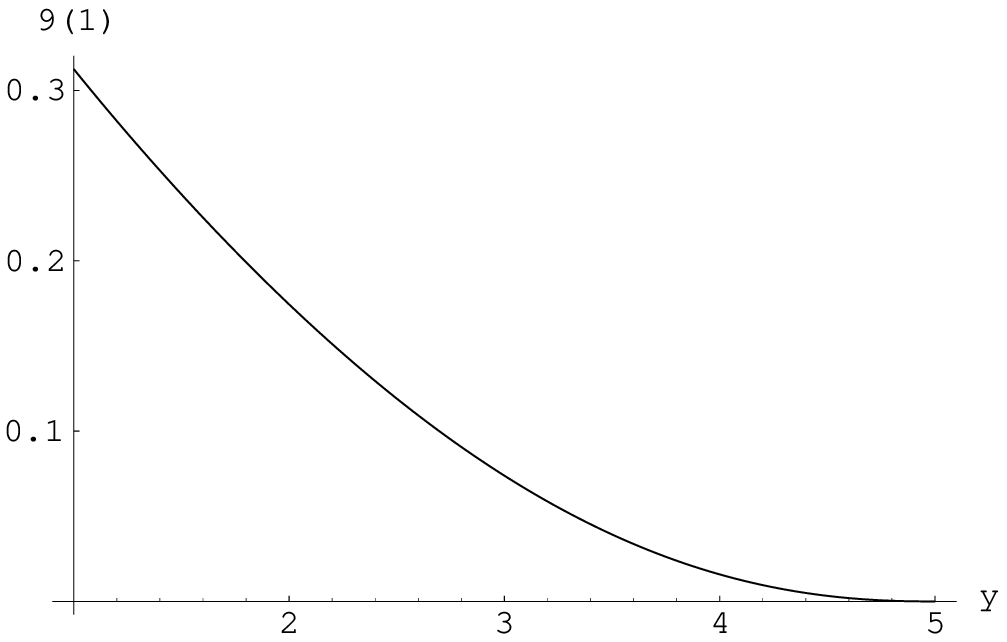}
\qquad
\epsfxsize=45mm
\epsfysize=35mm
\epsfbox{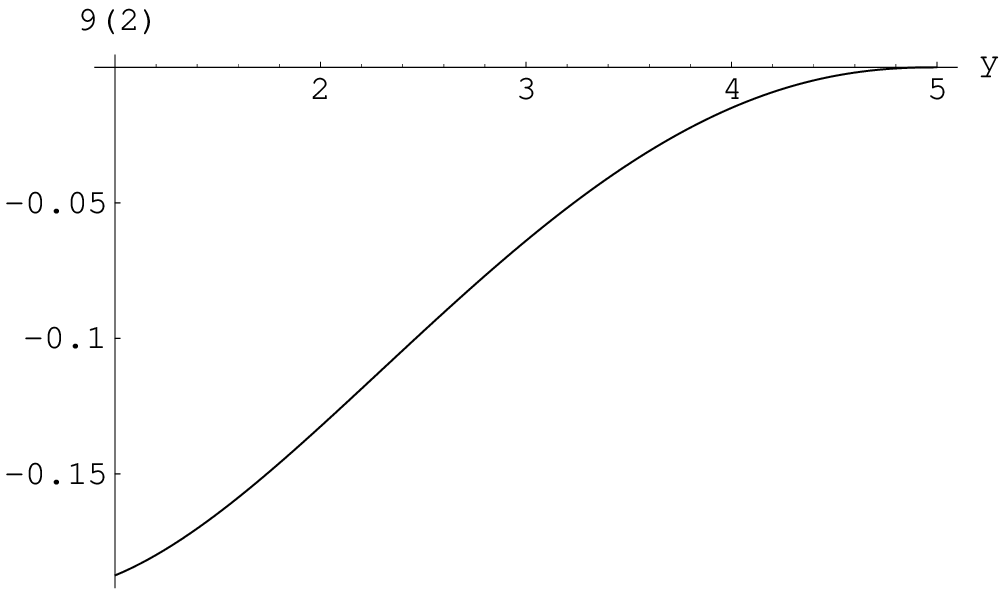}
\end{center}
\caption{{\small The effective potentials \Ref{Veffs} of the 
inert scalar fluctuation equation \Ref{eomR}, ($g\equ1/8$).}}
\la{FVeff}
\end{figure}

Without the $V_\Si$ contribution, the effective potential~\Ref{Veff}
may obviously be derived from a prepotential in the spirit of
supersymmetric quantum mechanics (cf.~\cite{DWoFre00}
for a more detailed discussion)
\be
\CV_\Si=\CU_\Si' + \CU_\Si^2 \;,
\la{susyqm}
\ee
with $\CU_\Si=\ft12A'(z)$. It has further been noted
in~\cite{DWoFre00} that the effective potentials for the active scalar
fluctuations in the most prominent exact five-dimensional
flows~\cite{GPPZ00,BraSfe99,FGPW00} may again be recast into the
form~\Ref{susyqm} with modified prepotentials $\CU_\Si$ (though no
general prescription has emerged). The absence of tachyonic
fluctuations is then manifest.

We find the same result for all effective potentials~\Ref{Veffs};
using \Ref{Ar}, they may be obtained from prepotentials
as~\Ref{susyqm} with
\ba
\CU_{\bf 1} &=& -\ft14\,g e^A\,(y-1)(7y-3) 
\qquad\mbox{or}\quad 
\CU_{\bf 1} ~=~ \ft14\,g e^A\,(y-1)(y+11) 
\non
\CU_{\bf 9{\rm (1)}} &=& -\ft14\,g e^A\,(y-1)(y-5) 
\qquad\mbox{or}\quad 
\CU_{\bf 9{\rm (1)}} ~=~ -\ft14\,g e^A\,(y-1)(y+11) 
\non
\CU_{\bf 9{\rm (2)}} &=& \ft14\,g e^A\,(y-1)(y-5) 
\non
\CU_{\bf 4{\rm (1)}} &=&  \ft14\,g e^A\,(y-1)(y+3) 
\non
\CU_{\bf 4{\rm (2)}} &=&  -\ft14\,g e^A\,(y-1)(y+3) \;.
\la{Ui}
\ea
{}From~\Ref{susyqm}, it follows immediately that a solution to
\Ref{eomR} which is normalizable as
\be
\int_0^\infty |R^\Si(z)|^2  \,dz ~<~ \infty \;,
\la{norm}
\ee
implies $p\ge0$, i.e.\ despite their unnerving shapes
(cf.~Figure~\ref{FVeff}), all the potentials~\Ref{Veffs} have positive
spectrum. Note that~\Ref{norm} corresponds to the norm $e^{A(z)}dz$
for the original scalar fluctuations~$\phi^\Si$, see~\cite{DFGK00} for
a detailed discussion on the proper choice of the norm.

The fact that the prepotentials 
$\CU_{\bf 9{\rm (1)}}$, $\CU_{\bf 9{\rm (2)}}$ and
$\CU_{\bf 4{\rm (1)}}$, $\CU_{\bf 4{\rm (2)}}$ 
respectively just differ by a sign
means that they are superpartners in the sense of supersymmetric
quantum mechanics. Specifically, the corresponding solutions
of~\Ref{eomR} may be related by
\be
R^{\bf 9{\rm (2)}} ~=~ 
(\dd_z + \CU_{\bf 9{\rm (2)}})\,R^{\bf 9{\rm (1)}} \;,
\la{sqmex}
\ee
and so on. This may be viewed as a consequence of the fact that
$\phi^{\bf 9{\rm (1)}}$ and $\phi^{\bf 9{\rm (2)}}$ are part of the
same supermultiplet under the remaining $N\equ(1,1)$ symmetry which
governs the flow. It is more surprising that even the prepotential
$\CU_{\bf 1}$ appears as a superpartner of $\CU_{\bf 9{\rm (1)}}$,
i.e.\ the corresponding potentials have the same spectrum, and their
solutions may likewise be mapped onto each other. The inert scalar
fluctuation equations with effective potentials~\Ref{Veffs} may thus
be reduced to just two independent differential equations.

Consider first $\CV_{\bf 9{\rm (1)}}$. One of its (non-normalizable)
zero modes (i.e.\ solutions of~\Ref{eomR} with $p\equ0$) may be found
from
\be
\dd_z R^{\bf 9{\rm (1)}}_0 ~=~ 
ge^{A}\,(y\!-\!5)\,(y\!-\!1)\,(y\!+\!1)\,\dd_y R^{\bf 9{\rm (1)}}_0 
{}~\stackrel{!}{\equiv}~ 
\CU_{\bf 9{\rm (1)}}\,R^{\bf 9{\rm (1)}}_0 \;,
\ee
which has the solution
\be
R_0^{\bf 9{\rm (1)}} ~=~ c_0 \,(1+y)^{-1/4} \;.
\ee
Dividing out the zero mode from the general solution 
$R^{\bf 9{\rm (1)}} =R^{\bf 9{\rm (1)}}_0 \chi$
leads to
\be
\tilde{p}^2\,  (1 + t^3 )\, \chi(t) - 2\, \chi'(t) 
+  t \,\chi''(t) ~=~ 0 \;, 
\la{ode1}
\ee
with $\tilde{p}=2^{-5/6}3^{-1/2}\,p$,
$y=\frac{5t^3-1}{1+t^3}$. Exploiting the supersymmetric quantum
mechanics structure~\Ref{sqmex}, the fluctuation equations for
$\CV_{\bf 1}$ and $\CV_{\bf 9{\rm (2)}}$ are reduced to the same
differential equation. The second differential equation is obtained by
similar considerations for $\CV_{\bf 4{\rm (2)}}$. A zero mode to this
potential is given by
\be
R^{\bf 4{\rm (2)}}_0 ~=~ c_0 \,\frac{(1+y)^{1/12}}{(y-5)^{1/3}} \;,
\ee
and dividing it out $R^{\bf 4{\rm (2)}}=R^{\bf 4{\rm (2)}}_0 \chi$
leads to
\be
\tilde{p}^2\,  (1 + t^3 )\, \chi(t) + 2\, \chi'(t) 
+  t \,\chi''(t) ~=~ 0 \;,
\la{ode2}
\ee
with $\tilde{p}$, $t$ defined as above. The two ordinary differential
equations~\Ref{ode1}, \Ref{ode2} thus contain the entire dynamics of
the inert scalar fluctuations.

\subsection{Vector/Scalar mixing}

Let us now consider the sectors in which scalars and vectors are
related by the local gauge symmetry. According to~\Ref{Sinv}, under
$SO(4)_{\rm inv}$ these are two copies of the ${\bf (1,3)}+{\bf
(3,1)}$ which we will denote by $\{ {\bf 6{\rm (1)}}, {\bf 6{\rm (2)}}
\}$ in the following. The Lagrangian in this sector is obtained as in
the preceding section from evaluating~\Ref{L} with an
ansatz~\Ref{infsc} where the sum now runs over the corresponding
representations $\Si\in\{{\bf 6{\rm (1)}}, {\bf 6{\rm (2)}}\}$,
including additional contributions from the Chern-Simons and the
kinetic scalar term. Again, somewhat miraculously the effective
potential may be diagonalized with $q$-independent eigenvectors, such
that the resulting Lagrangian is given by
\be
\CL ~=~ \textstyle{\sum_\SBi}\: \CL_{B^\SBi,\phi^\SBi}
\;,\qquad \SBi \in \{ {\bf 6{\rm (1)}}, {\bf 6{\rm (2)}} \} \;,
\ee
with
\ba
\CL_{B^\SBi,\phi^\SBi} &=& 
\ft14\sqrt{G}G^{\mu\nu}\,\dd_\mu\phi^\SBi\,\dd_\nu\phi^\SBi
- g^2\sqrt{G}\,V_\SBi(q)\,\phi^\SBi\,\phi^\SBi
+ \ft12\sqrt{G}\,\Ge^{\rho\mu\nu}\,B^\SBi_k\,F^\SBi_{\mu\nu} \non
&&{}
+ g\sqrt{G}\,G^{\mu\nu} B^\SBi_\mu (  Z_\SBi(q) \dd_\nu\phi^\SBi - 
\phi^\SBi \dd_\nu Z_\SBi(q))
+ g^2\sqrt{G} G^{\mu\nu} \,B^\SBi_\mu B^\SBi_\nu \,Z_\SBi(q)^2 \;,
\la{Lvc}
\ea
and their equations of motion decouple. The effective mass terms
$V_\SBi(q)$ and $Z_\SBi(q)$ for scalars and vector fields,
respectively, are related by
\be
V_\SBi ~=~ 
-\frac1{4 \sqrt{G}\,Z_\SBi}\,
\dd_\mu(\sqrt{G}G^{\mu\nu}\dd_\nu Z_\SBi) \;,
\la{VZ}
\ee
with the metric \Ref{Ar}, \Ref{conmet}, and explicitly given by
\ba
V_{\bf 6{\rm (1)}} &=& \ft1{16}\,
(y+1)(y-5)(17 + 4\,y + 3\,y^2)
\non
V_{\bf 6{\rm (2)}} &=& \ft1{4}\,
(y-5)( 5 + 5\,y + 2\,y^3) \;,
\la{ViB}
\ea
and 
\be
Z_{\bf 6{\rm (1)}} ~=~ \sqrt{2(y-1)}
\;,\qquad
Z_{\bf 6{\rm (2)}} ~=~ \sqrt{y^2-1} \;.
\la{ZZ}
\ee
The linearized local gauge symmetry~\Ref{gaugeG} acts as
\be
\Gd B^\SBi_\mu ~\ra~ \dd_\mu\,\GL^\SBi \;,\qquad
\Gd \phi^\SBi ~\ra ~ -2g Z_\SBi(q) \GL^\SBi \;,
\la{gaugelin}
\ee
which leaves \Ref{Lvc} invariant, as may be explicitly checked making
use of~\Ref{VZ}. The equations of motion obtained from variation
of~\Ref{Lvc} give the duality equations relating vector and scalar
fields
\be
\Ge^{\mu\nu\rho}\,F^\SBi_{\mu\nu} ~=~ 
-2g\sqrt{G}\, Z_\SBi(q)^2\,G^{\rho\nu} \Big(B^\SBi_\nu +
\frac1{2g}\,\dd_\nu( Z_\SBi(q)^{-1}\,\phi^\SBi ) \Big) \;,
\la{dualq}
\ee
and the scalar equation of motion which may be obtained from the
Bianchi identities implied by \Ref{dualq}. We may fix
the gauge \Ref{gaugelin} by setting $\phi^\SBi$ to zero, but will
prefer to equivalently work with the gauge invariant object
\be
\CB^\SBi_\mu ~\equiv~ B^\SBi_\mu + 
\frac1{2g}\,\dd_\mu( Z_\SBi^{-1}\phi^\SBi ) \;.
\ee
The duality equation \Ref{dualq} then simply takes the form
\be
\Ge^{\rho\mu\nu}\,F^\SBi_{\mu\nu} ~=~ 
-2g\sqrt{G}\, Z_\SBi^2\,G^{\rho\nu} \CB^\SBi_\nu \;,
\ee
while the scalar equation of motion gives the current conservation
\be
\dd_\mu\,(\sqrt{G} Z_\SBi^2 G^{\mu\nu} \CB^\SBi_\nu ) ~=~ 0 \,.
\ee
Combining these two equations and assuming a metric of conformal form
\Ref{conmet}, leads after some computation to 
\ba
(G^{\mu\nu}\dd_\mu\dd_\nu)\,\CB^\SBi_\rho &=& 
-g^2 Z_\SBi^4\,\CB^\SBi_\rho
-(\dd_\rho + 2\dd_\rho A) \Big(\CB^\SBi_\mu
G^{\mu\nu}\dd_\nu(A+2\ln Z_\SBi) \Big) 
\non
&&{}
- F^\SBi_{\rho\mu}\,G^{\mu\nu}\dd_\nu (A+2\ln Z_\SBi) 
\la{eomvec}\;.
\ea
With the ansatz $\CB^\SBi_z = e^{-i (p\cdot x)} e^{-A(z)/2}
Z_\SBi^{-1} {b}\,^\SBi_z(z)\,$, the corresponding component of the
vector equations of motion \Ref{eomvec} turns into
\be
(-\dd_z^2 + \CV_{\SBi}\,)\, b\,^\SBi_z ~=~ p^2\, b\,^\SBi_z \;,
\ee
where the effective potentials are found after some computation to be
\be
\CV_{\SBi} ~=~ g^2 e^{2A}\,\Big(
Z_\SBi^4 + \frac12\,(\ln'Z_\SBi)^2 W'{}^2 
-\frac{W'}{4Z_\SBi}\,(W'Z_\SBi')' 
+ \frac12\,W'{}^2 -W^2
\Big)
\;,
\ee
with primes denoting derivatives with respect to~$q$. Inserting the
superpotential \Ref{W} and the scalar functions from \Ref{ZZ} then
yields:
\ba
\CV_{{\bf 6{\rm (1)}}} &=& 
-\ft5{16}\,g^{2} e^{2A}\,\,(y-1)^2\,(y-5)(y+3)
\non
\CV_{{\bf 6{\rm (2)}}} &=& 
\ft1{16}\,g^{2} e^{2A}\,(y-1)^2\,( 167 + 34\,y + 7\,y^2 ) \;.
\la{Veffvc}
\ea
Surprisingly, these effective potentials precisely coincide with the
effective potentials for the scalars $\CV_{\bf 9{\rm (1)}}$ and
$\CV_{\bf 1}$, respectively!  In particular, \Ref{Ui} shows that
$\CV_{{\bf 6{\rm (1)}}}$ and $\CV_{{\bf 6{\rm (2)}}}$ admit
prepotentials and are superpartners in the sense of supersymmetric
quantum mechanics. Following the analysis of the last section, the
equations of motion for the longitudinal vector components $\CB_z$
thus may again be reduced to the ordinary differential
equation~\Ref{ode1}. The transverse components $\CB_0$, $\CB_1$ are
finally obtained from the duality equation \Ref{dualq} as explicit
functions of $\CB_z$ and its derivatives. This is in agreement with
the fact that the three-dimensional vectors carry the same number of
degrees of freedom as the scalar fields.

This finishes our computation of fluctuation equations in the background
\Ref{kink}. As discussed above, these equations in principle allow us to
compute 2-point functions throughout the flow. We leave this endeavor
to future work.

\section{Conclusion}

We have found an analytic domain wall solution in three-dimensional
gauged supergravity which describes an $N\equ(1,1)$ supersymmetric RG
flow between conformal fixed points of a two-dimensional quantum field
theory. It is driven by a relevant operator of conformal dimension
$\Delta\equ\ft32$. We have computed counterterms to the Lagrangian,
and the 1-point functions for inert scalars in the presence of
sources. Finally, we have derived the fluctuation equations for inert
scalars and vector fields which reduce to the second order
differential equations~\Ref{ode1}, and~\Ref{ode2}.

While these differential equations appear fairly simple, they are not
of hypergeometric type as were those encountered in higher-dimensional
examples previously treated in the literature. So far we have not
found analytic solutions to these equations. They would immediately
yield the 2-point correlation functions,
cf.~\cite{DWoFre00,BWFP01,BiFrSk01,Muck01,BiFrSk01a}. However, this
information may in principle also be extracted numerically
from~\Ref{ode1}, \Ref{ode2}. We stress that the crucial step in the
whole analysis was finding an analytic kink solution, whereas a purely
numerical description of this flow would not be
sufficient. In contrast, an analytic solution to the
fluctuation equations would certainly be helpful and of interest, but
is not required for the final computation of correlation functions.
\footnote{General properties of holographic CFT/CFT flows
have recently been studied in \cite{MarMie02}. See also
\cite{PorSta99} for earlier work on model-independent statements on
holographic RG flows.}

In the course of our discussion, several ``miracles'' passed before
our eyes.  The ratio of central charges in the $\alpha\!=\!1$ case
comes out rational, just like in the FGPW flow of 3+1 boundary theory
\cite{FGPW99} --- despite being given by a square root formula. For
the extremum we study, even the conformal dimensions in the infrared
come out rational, something which did not happen in the FGPW
flow. Note that the same phenomena occur for the nonsupersymmetric but
stable extremum $c)$ from table~\ref{extrema}; it has rational central
charge as well as rational conformal dimensions, cf.~table~V, and
might deserve further study.  

Further, the Lagrangian for the inert scalars and the vector fields
was found to be diagonalizable, such that the equations of motion of
these fields led to uncoupled second order differential equations. All
of these equations admitted simple prepotentials in the sense of
supersymmetric quantum mechanics. Even more surprising, we found that
despite being scattered on several distinct supermultiplets, all
fluctuation equations of motion could be reduced to just two quite
innocuous equations~\Ref{ode1}, \Ref{ode2}. There seems to be no
direct reason why the effective potential for the singlet scalar
$\CV_{\bf 1}$, as well as the effective potential for the longitudinal
vector fields $\CV_{{\bf 6}(1)}$, would turn out to be nothing but the
superpartners of $\CV_{{\bf 9}(1)}$. From a pure supergravity
perspective, the occurrences of rational central charges as well as
rational conformal dimensions around the supersymmetric extremum may
seem like black magic; they presumably admit rational explanation from
the point of view of the holographic renormalization group --- lending
further support to the latter as an extension of the AdS/CFT
correspondence.

As this paper draws to a close, let us give some general ideas of
future directions of study. As an immediate application, the 2-point
functions of operators dual to inert scalars and vectors should be
extracted from the fluctuation equations given here.  Two-point
functions of the operator $\CO_q$ dual to the active scalar $q$ and
the stress-energy tensor require some additional calculations but are
in principle straightforward to obtain using the formalism of
holographic renormalization.  This should then settle, among other
things, whether the flow presented here is truly a mass deformation.
As we emphasized in the introduction, the relatively advanced
understanding of 1+1 conformal field theories can then be exploited so
that quantities like the $C$ function can actually be directly
computed in the dual field theory, at least in principle; of course
this task is not entirely trivial when the theory is strongly coupled.

Further, most of our discussion qualitatively applies to general
values of $\alpha$ (the ratio of coupling constants in the gauged
supergravity, or the ratio of $SU(2)$ levels in the boundary CFT), and
other values than our main example $\alpha=1$ are interesting to
study. We picked $\alpha=1$ as an example partly due to technical
simplifications.  By turning on scalars at other values of $\alpha$,
we can study RG flows from large $N=4$ superconformal field theories,
driven by operators of dimension $\Delta_+=1+\ft1{1+\alpha}$, which
for small $\alpha$ approaches marginality. In fact, this limit may be
of particular interest; in the double D1-D5 system (see the
introduction), $\alpha \rightarrow 0$ corresponds to one of the two
D1-D5 system decoupling, a limit with subtleties of its own.  We hope
that eventually, RG flows in three-dimensional gauged supergravity
theories will shed some light on certain aspects of the quantum
mechanics of nonextremal black holes.

Another very interesting topic is a chiral breaking of supersymmetry,
e.g.\ $N\equ(4,4)$ breaking to $N\equ(4,0)$ in these models. As
discussed in the main text, this cannot be accomplished by flows of
the type studied here, which turn on nothing but scalar fields. In
this context it would be most interesting to understand, within the
framework presented here, the higher-dimensional $N\equ(4,0)$ solution
recently constructed in
\cite{MorTri01}.

Finally, let us turn the spotlight to a related scene of interest: the
maximal three-dimensional gauged supergravity, with 32 supercharges,
that was constructed in~\cite{NicSam01,NicSam01a}. That theory enjoys
even closer analogy to the maximal theories used in five dimensions,
and while being more involved technically (for instance, scalars
parametrizing the exceptional coset manifold $E_{8(8)}/SO(16)$ rather
than the orthogonal ones encountered here) it exhibits several very
intriguing features.

\section*{Acknowledgements}

We gratefully acknowledge useful discussions with C.~Bachas, P.~Bain,
J.~de~Boer, R.~McNees, F.~Morales, I.~Runkel, K.~Skenderis,
B.~Sundborg, M.~Trigiante, P.~Vanhove, F.~Zamora, and we especially
thank M.~Bianchi and M.~Haack for useful comments on early drafts of
this paper. M.B. is supported by a Marie Curie Fellowship, contract
number HPMF-CT-2001-01311. This work is partly supported by EU
contract HPRN-CT-2000-00122.

\begin{appendix}
\section{Scalar potential and stable extrema}
\la{AppExt}

In this appendix, we give an explicit parametrization of the scalar
potential~\Ref{potential} for $n\equ4$ matter multiplets and collect
some of its stable extrema together with the spectrum around these
points. 

As discussed in the main text, a naive counting suggests that this
potential is a function of 14 variables. To begin with, the matrix
$\Mat$ is an element of $SO(8,4)$. The freedom of right multiplication
by compact $SO(8)\cro SO(4)$ matrices, corresponding to the coset
structure of the scalar manifold, may be fixed by bringing $\Mat$ into
the form
\be
\Mat ~=~ \left( 
\begin{array}{cc}
\sqrt{I_4 + Y^T Y} & Y^T \\
Y & \sqrt{I_8 + Y Y^T}
\end{array}
\right) \;,
\la{gf}
\ee
with an $8\times4$ matrix $Y$. In this representation system of the
coset manifold, the local $SO(4)\cro SO(4)$ invariance acts by
conjugation on $\Mat$, i.e.\ as
\be
Y ~\ra~ \GL_1 Y \;, \qquad \GL_1\in  SO(4)\cro SO(4) \;.
\la{l1}
\ee
The global $SO(4)$ symmetry rotating the matter multiplets likewise
acts by conjugation on $\Mat$ such that
\be
Y ~\ra~ Y \GL_2 \;, \qquad \GL_2\in  SO(4)\;.
\la{l2}
\ee
Fixing~\Ref{l2} in the singular value decomposition of $Y$, allows to
bring it into the form
\be
Y ~=~ S\,D_2 \;,\qquad S\in SO(8)\;,\quad
D_2 = \left( 
\begin{array}{c}
\diag(q_1,q_2,q_3,q_4) \\
0
\end{array}
\right)
\;.
\ee
The $SO(8)$ matrix $S$ may be decomposed into $S= T_1D\,T_2$ with
block matrices $T_1, T_2\in SO(4)\cro SO(4)$, and $D=((\cos D_1,
\sin D_1),(-\sin D_1, \cos D_1))$, with a $4\cro4$ diagonal
matrix~$D_1$. Fixing~\Ref{l1} to absorb $T_1$, and changing the
representation system~\Ref{gf} by a final $SO(4)$ rotation from the
right hand side to absorb one of the $SO(4)$ blocks from $T_2$, we may
eventually bring $\Mat$ into the form
\ba
\Mat &=&
\left( 
\begin{array}{ccc}
I_4&&
\\
&\cos D_1&\sin D_1
\\
&-\sin D_1& \cos D_1
\end{array}
\right)
\left( 
\begin{array}{ccc}
I_4&&
\\
&T&
\\
&& I_4
\end{array}
\right)
\left( 
\begin{array}{ccc}
\cosh D_2 & \sinh D_2 &
\\
\sinh D_2 & \cosh D_2 &
\\
&&I_4
\end{array}
\right)
\;,
\la{par}\\[2ex]
&& D_1=\diag(p_1,p_2,p_3,p_4)\;,\qquad
D_2=\diag(q_1,q_2,q_3,q_4)\;,\quad
T\in SO(4) \;.
\nn
\ea
This contains precisely $14$ parameters; all the redundancies are
fixed. Inserting~\Ref{par} back into~\Ref{potential} yields the scalar
potential $V$ as explicit function of these 14~variables. We have
numerically found some extrema on this space, all of which exhibit
$T\equ I_4$, i.e.\ live in the truncation~\Ref{matL}. The extremal
points preserving some remaining symmetry have been listed in
table~\ref{extrema}; $b)$ and $d)$ preserve $N\equ(1,1)$
supersymmetry, while $c)$ has an unbroken $SO(4)$ diagonal of the
gauge group~\Ref{gaugeG}. The spectrum of physical fields around $d)$
has been given in table~\ref{specL0} and further exploited in the main
text. For completeness, we give here the spectra around $b)$ and $c)$,
computed in analogy to table~\ref{specL0}. The former one which
decomposes into $N=(1,1)$ supermultiplets is listed in
table~\ref{Extb}. The latter is organized by the unbroken $SO(4)$ and
presented in table~V, it corresponds to a nonsupersymmetric
two-dimensional CFT with rational central charge $c_{\rm IR}/c_{\rm
UV} = 2/3$.

\begin{table}[htb]
\centering
\begin{tabular}{|c||l|c|c|} \hline
$SU(2)_{\rm inv}$ & fields & $m^2\adsr^2$ & $(h,\bar{h})$ \\
\hline\hline
{\bf 1} & scalars & 
$-\ft12, \ft{11+8\sqrt{2}}{6+4\sqrt{2}}$ & 
$(\ft12\pls\ft1{2\sqrt{2}},\ft12\pls\ft1{2\sqrt{2}})+
(1\pls\ft1{2\sqrt{2}},1\pls\ft1{2\sqrt{2}})$ \\
\cline{2-4}
&fermions & $\ft34+\ft1{\sqrt{2}}$ & 
$(\ft12\pls\ft1{2\sqrt{2}},1\pls\ft1{2\sqrt{2}})+
(1\pls\ft1{2\sqrt{2}},\ft12\pls\ft1{2\sqrt{2}})$ \\
\hline 
\hline 
{\bf 1} & scalars & 
$1, \ft{2(7+5\sqrt{2})}{3+2\sqrt{2}}$ & 
$(\ft12\pls\ft12\sqrt{2},\ft12\pls\ft12\sqrt{2})+
(1\pls\ft12\sqrt{2},1\pls\ft12\sqrt{2})$ \\
\cline{2-4}
&fermions & $\ft94+\sqrt{2}$ & 
$(\ft12\pls\ft12\sqrt{2},1\pls\ft12\sqrt{2})+
(1\pls\ft12\sqrt{2},\ft12\pls\ft12\sqrt{2})$ \\
\hline 
\hline 
{\bf 3} & scalars & 
$-1,0$ & $(1,1)+(\ft12,\ft12)$ \\
\cline{2-4}
&fermions & $\ft14$ & $(1,\ft12)+(\ft12,1)$ \\
\hline 
\hline 
{\bf 3} & fermions & $\ft34+\ft1{\sqrt{2}}$ & 
$(\ft12\pls\ft1{2\sqrt{2}},1\pls\ft1{2\sqrt{2}})$ \\
\cline{2-4}
&vectors & $\ft12,\ft32+\sqrt{2}$ & 
$(\ft1{2\sqrt{2}},1\pls\ft1{2\sqrt{2}})+
(\ft12\pls\ft1{2\sqrt{2}},\ft32\pls\ft1{2\sqrt{2}})$ \\
\cline{2-4}
&gravitini & $\ft34+\ft1{\sqrt{2}}$ & 
$(\ft1{2\sqrt{2}},\ft32\pls\ft1{2\sqrt{2}})$ \\
\hline 
\hline 
{\bf 3} & fermions & $\ft34+\ft1{\sqrt{2}}$ & 
$(1\pls\ft1{2\sqrt{2}},\ft12\pls\ft1{2\sqrt{2}})$ \\
\cline{2-4}
&vectors & $\ft12,\ft32+\sqrt{2}$ & 
$(1\pls\ft1{2\sqrt{2}},\ft1{2\sqrt{2}})+
(\ft32\pls\ft1{2\sqrt{2}},\ft12\pls\ft1{2\sqrt{2}})$ \\
\cline{2-4}
&gravitini & $\ft34+\ft1{\sqrt{2}}$ & 
$(\ft32\pls\ft1{2\sqrt{2}},\ft1{2\sqrt{2}})$ \\
\hline 
\hline 
{\bf 5} & scalars & 
$-\ft12, -\ft{5+4\sqrt{2}}{6+4\sqrt{2}}$ & 
$(\ft1{2\sqrt{2}},\ft1{2\sqrt{2}})+
(\ft12\pls\ft1{2\sqrt{2}},\ft12\pls\ft1{2\sqrt{2}})$ \\
\cline{2-4}
&fermions & $\ft34-\ft1{\sqrt{2}}$ & 
$(\ft1{2\sqrt{2}},\ft12\pls\ft1{2\sqrt{2}})+
(\ft12\pls\ft1{2\sqrt{2}},\ft1{2\sqrt{2}})$ \\
\hline 
\end{tabular}
\caption{\small Spectrum around extremum $b)$ (Table~\ref{extrema})
in $N=(1,1)$ supermultiplets}
\label{Extb}
\end{table}

In addition, we have found some extrema which do not preserve any rest
symmetry, but which are stable extremal points of the full
potential~\Ref{potential}, i.e.\ all $8n$ scalars satisfy the
Breitenlohner-Freedman bound~\cite{BreFre82}. The corresponding
boundary theories are non-supersymmetric CFTs with irrational central
charge. In coordinates~\Ref{pqxy}, these extremal points are located
at
\begin{itemize}

\item
\be
x_1=x_2=-y_1=-y_2=1\;,\qquad
x_3=-x_4=y_3=y_4=\sqrt{3} \;,
\la{Ex3} 
\ee
$\qquad\quad$ with $c_{\rm IR}/c_{\rm UV} ~=~ 
1/\sqrt{7}  ~\approx~ 0.3790\,$.

\item
\be
x_1=x_2=x_3=x_4=-y_1=y_2=y_3=y_4=
\ft{\sqrt{2+\sqrt{13}}}{\sqrt{3}} \;,
\la{Ex7} 
\ee
$\qquad\quad$ with $c_{\rm IR}/c_{\rm UV} ~= ~
\ft{3\sqrt{3}}{\sqrt{97+26\sqrt{13}}}~\approx~ 0.3762 \,$.

\item\be
x_1=-y_1=\ft{\sqrt{1+2\sqrt{5}}}{\sqrt{2}} \;,\qquad
x_2=x_3=x_4=y_2=y_3=y_4=
\ft{\sqrt{1+\sqrt{5}}}{\sqrt{2}} \;,
\la{Ex8} 
\ee
$\qquad\quad$ with $c_{\rm IR}/c_{\rm UV} ~= ~
\ft{4}{\sqrt{57 + 25 \sqrt{5}}} ~\approx~ 0.3765 \,$.

\end{itemize}

It may be noted as an amusing coincidence that their central charges
lie within a range of less than one percent. The physical spectra
around these extremal points are collected in tables~VI--VIII. We have
not included the fields which become massless due to the (super) Higgs
effect.

\begin{minipage}[htbp]{67.5mm}
\centering

\vspace*{28mm}

\begin{tabular}{|l|c|c|} \hline
fields & $\GD$ & $\#$  \\
\hline\hline
scalars & $\ft{10}{3}$ & 1 \\
\cline{2-3}
&$1\pm\ft23$& $16$ \\
\cline{2-3}
&$1\pm\ft13$& $9$ \\
\hline
fermions & $\ft{7}{6}$ & 24  \\
\hline
vectors & $\ft{7}{3}$ & 6 \\
\hline
gravitini & $\ft{11}{6}$ & 8 \\
\hline
\end{tabular} 

\medskip
Table V: {\small Spectrum around $c)$ (Table~\ref{extrema})}

\vspace*{15mm}

\centering
\begin{tabular}{|l|c|c|} \hline 
fields & $\GD$ & $\#$ \\
\hline\hline
scalars & $1\pm\ft{2}{\sqrt{7}}$& $6$ \\
\cline{2-3}
&$1\pm\ft{\sqrt{3}}{\sqrt{7}}$ & 4 \\
\cline{2-3}
&$1\pm1$& $4$ \\
\cline{2-3}
&$1+\ft{2\sqrt{3}}{\sqrt{7}}$& $2$ \\
\cline{2-3}
&$1+\ft{3\sqrt{2}+2}{\sqrt{7}}$& $1$ \\
\cline{2-3}
&$1\pm\ft{3\sqrt{2}-2}{\sqrt{7}}$& $1$ \\
\cline{2-3}
&$1\pm \ft{\sqrt{22-4\sqrt{21}}}{\sqrt{7}}$& $1$ \\
\cline{2-3}
&$1+\ft{\sqrt{22+4\sqrt{21}}}{\sqrt{7}}$& $1$ \\
\hline
fermions & $1$ & $12$  \\
\cline{2-3}
& $3$ & $4$  \\
\cline{2-3}
& $1+\ft{\sqrt{23 - 3 \sqrt{57}}}{\sqrt{14}}$ & $4$ \\
\cline{2-3}
& $1+\ft{\sqrt{23 + 3 \sqrt{57}}}{\sqrt{14}}$ & $4$ \\
\hline
vectors & $1+\ft4{\sqrt{7}}$ & 8 \\
\cline{2-3}
&$1+\ft3{\sqrt{7}}$ & 2 \\
\cline{2-3}
&$1+\sqrt{3}$ & 2 \\
\hline
gravitini & $1+\ft{2\sqrt{3}}{\sqrt{7}}$ & 4 \\
\cline{2-3}
& $2$ & $4$ \\
\hline
\end{tabular} 

\medskip
Table~VI: {\small Spectrum around~\Ref{Ex3}}
\end{minipage}
\begin{minipage}[htbp]{85mm}
\centering
\begin{tabular}{|l|c|c|} \hline 
fields & $\GD$ & $\#$ \\
\hline\hline
scalars & $1\pm\ft{\sqrt{2(3\sqrt{13}-5)}}{\sqrt{23}}$& $9$ \\
\cline{2-3}
&$1+\ft{\sqrt{2(17-\sqrt{13})}}{\sqrt{23}}$& $9$ \\
\cline{2-3}
&$1+\ft{\sqrt{2(31+9\sqrt{13})}}{\sqrt{23}}$& $1$ \\
\cline{2-3}
&$1+\ft{\sqrt{2(53+5\sqrt{13})}}{\sqrt{23}}$& $1$ \\
\hline
fermions & $1$ & $16$  \\
\cline{2-3}
& $1+\ft{\sqrt{71+31\sqrt{13}}}{\sqrt{46}}$ & $8$ \\
\hline
vectors 
&$1+\ft{\sqrt{223+62\sqrt{13}}}{\sqrt{97+26\sqrt{13}}}$ & 12 \\
\hline
gravitini & 
$1+\ft{\sqrt{31+9\sqrt{13}}}{\sqrt{46}}$ & 8 \\
\hline
\end{tabular} 

\medskip
Table VII: {\small Spectrum around~\Ref{Ex7}}

\bigskip
\bigskip
\medskip

\centering
\begin{tabular}{|l|c|c|} \hline 
fields & $\GD$ & $\#$ \\
\hline\hline
scalars & $1\pm\ft{\sqrt{-141+70\sqrt{5}}}{\sqrt{31}}$& $8$ \\
\cline{2-3}
& $1+\ft{\sqrt{-139+80\sqrt{5}}}{\sqrt{31}}$& $5$ \\
\cline{2-3}&$1\pm1$& $3$ \\
\cline{2-3}
&$1+\ft{\sqrt{-20+55\sqrt{5}+\sqrt{24286-7530\sqrt{5}}}}{\sqrt{31}}$&
 $1$ \\
\cline{2-3}
&$1\pm\ft{\sqrt{-20+55\sqrt{5}-\sqrt{24286-7530\sqrt{5}}}}{\sqrt{31}}$&
 $1$ \\
\cline{2-3}
&$1+\ft{\sqrt{65+15\sqrt{5}+\sqrt{22601-7740\sqrt{5}}}}{\sqrt{31}}$&
 $1$ \\
\cline{2-3}
&$1\pm\ft{\sqrt{65+15\sqrt{5}-\sqrt{22601-7740\sqrt{5}}}}{\sqrt{31}}$&
 $1$ \\
\hline
fermions & $ 1 + \ft{\sqrt{-313 + 140\sqrt{5}}}{\sqrt{62}} $ & $16$  \\
\cline{2-3}
& $1 + \ft{\sqrt{5(-13 + 28\sqrt{5})}}{\sqrt{62}} $ & $6$  \\
\cline{2-3}
& $1 + \ft{\sqrt{47 + 80\sqrt{5}}}{\sqrt{62}} $ & $2$ \\\hline
vectors & $1 + 
\ft{\sqrt{6(23 + 11\sqrt{5})}}{\sqrt{57 + 25\sqrt{5}}}$ & 6 \\
\cline{2-3}
&$1 + 
\ft{\sqrt{2(59 + 27\sqrt{5})}}{\sqrt{57 + 25\sqrt{5}}}$ & 6 \\
\hline
gravitini & $1 + \ft{\sqrt{5(7 + 4\sqrt{5})}}{\sqrt{62}}$ & 6 \\
\cline{2-3}
&  $1 + \ft{\sqrt{-77 + 80\sqrt{5}}}{\sqrt{62}}$ & 2 \\
\hline
\end{tabular} 

\medskip
Table~VIII: {\small Spectrum around~\Ref{Ex8}}

\end{minipage}

\end{appendix}

\newpage


\providecommand{\href}[2]{#2}\begingroup\raggedright\endgroup

\end{document}